\documentclass[twocolumn]{aastex631}

\graphicspath{{./}{figures/}}
\usepackage{amsmath}
\usepackage{upgreek}
\usepackage[T1]{fontenc}
\usepackage{booktabs}

\usepackage{CJK}

\newcommand\sref[1]{\hyperref[#1]{\S~\ref*{#1}}}
\newcommand\fref[1]{\hyperref[#1]{Fig.~\ref*{#1}}}
\newcommand\Eqref[1]{Eq.~(\hyperref[#1]{\ref*{#1}})}
\newcommand\eeqref[1]{Eq.~\hyperref[#1]{\ref*{#1}}}
\newcommand\tref[1]{\hyperref[#1]{Table~\ref*{#1}}}
\newcommand\aref[1]{\hyperref[#1]{Appendix~\ref*{#1}}}

\shorttitle{Magnetically Arrested Accretion and Black Hole Spin in a Semi-Analytic Model of Galaxy Formation}
\shortauthors{Porras-Valverde et al.}

\begin{document}
\begin{CJK}{UTF8}{mj}
\title{Magnetically Arrested Accretion and Black Hole Spin in a Semi-Analytic Model of Galaxy Formation}

\author[0000-0002-1996-0445]{Antonio J. Porras-Valverde}
\affiliation{Department of Astronomy, Yale University, Kline Tower, 266 Whitney Avenue, New Haven, CT 06511, USA}
\email{antonio.porras@yale.edu}

\author[0000-0001-5287-0452]{Angelo Ricarte}
\affiliation{Black Hole Initiative at Harvard University, 20 Garden Street, Cambridge, MA 02138, USA}
\affiliation{Center for Astrophysics $\vert$ Harvard \& Smithsonian, 60 Garden Street, Cambridge, MA 02138, USA}
\email{angelo.ricarte@cfa.harvard.edu}

\author[0000-0002-5554-8896]{Priyamvada Natarajan}
\affiliation{Department of Astronomy, Yale University, Kline Tower, 266 Whitney Avenue, New Haven, CT 06511, USA}
\affiliation{Department of Physics, Yale University, P.O. Box 208121, New Haven, CT 06520, USA}
\affiliation{Black Hole Initiative at Harvard University, 20 Garden Street, Cambridge, MA 02138, USA}
\email{priyamvada.natarajan@yale.edu}

\author[0000-0002-2858-9481]{Hyerin Cho (조혜린)}
\affiliation{TAPIR, Mailcode 350-17, California Institute of Technology, Pasadena, CA 91125, USA}
\affiliation{Center for Astrophysics $\vert$ Harvard \& Smithsonian, 60 Garden Street, Cambridge, MA 02138, USA}
\affiliation{Black Hole Initiative at Harvard University, 20 Garden Street, Cambridge, MA 02138, USA}

\author[0000-0002-1919-2730]{Ramesh Narayan}
\affiliation{Center for Astrophysics $\vert$ Harvard \& Smithsonian, 60 Garden Street, Cambridge, MA 02138, USA}
\affiliation{Black Hole Initiative at Harvard University, 20 Garden Street, Cambridge, MA 02138, USA}

\author[0000-0003-1598-0083]{Kung-Yi Su}
\affiliation{Department of Physics \& Astronomy and Center for Interdisciplinary Exploration and Research in Astrophysics(CIERA), Northwestern University, 1800 Sherman Ave, Evanston, IL 60201, USA}

\author[0000-0002-0393-7734]{Cora Prather} 
\affiliation{Black Hole Initiative at Harvard University, 20 Garden Street, Cambridge, MA 02138, USA}

\begin{abstract}\label{abstract}
We present {\sc Mad Sage}, a semi-analytic model of galaxy formation that couples black hole mass and spin evolution to gas accretion and AGN feedback under magnetically arrested disk conditions. Building on the {\sc Dark Sage} model, we incorporate GRMHD-calibrated prescriptions for spin evolution and jet power, spin-dependent radiative efficiencies, and remnant spins from BH-BH mergers. We compare accretion and feedback models, including coherent disk-fed accretion with mixed coherent and incoherent hot-halo accretion, with and without additional mechanical feedback. We find that incoherent accretion, together with Blandford-Znajek spin-down, drives massive BHs toward low spins, reducing jet efficiencies and weakening the quenching of massive galaxies. Reproducing the high-mass end of the stellar mass function requires either exclusively coherent accretion or an additional feedback component provided by ADAF outflows. Our models broadly reproduce observed BH-galaxy scaling relations, disk surface density profiles, and the AGN luminosity function at $z\sim 2-6$, although the spin models under-predict the number density of faint AGN at low redshift. At $z\lesssim 3$, predicted BH spins decrease with increasing BH mass, while at higher redshifts, the mass dependence weakens. Spin distributions peak at $|a_\star|\sim 0.5$-$0.7$, below current X-ray reflection spectroscopy measurements, and become bimodal for massive, quiescent BHs accessible with the next generation Event Horizon Telescope and the Black Hole EXplorer. We report that the predicted LISA event rates are nearly independent of the adopted accretion and spin models.
\end{abstract}


\section{Introduction}
\label{S:intro}

Supermassive black holes (SMBHs) reside at the centers of most massive galaxies and are now understood to play a critical role in regulating star formation through Active Galactic Nuclei (AGN) feedback \citep{KormendyRichstone1995, SilkRees1998, Fabian1999, Murray2005, Fabian2012, KormendyHo2013, Su+2020, Terrazas2020}. The mass evolution of SMBHs has been studied extensively through quasar luminosity functions, black hole (BH) mass functions, and scaling relations with host galaxy properties \citep{Magorrian1998, Tremaine2002, Haring2004, Graham2007, KormendyHo2013, Reines&Volonteri2015, Schutte2019, Greene2020, Sturm&Reines2024, Burke2025}. Yet astrophysical BHs are characterized by two physical parameters: mass and spin. While the mass-evolution of SMBHs has been the focus of considerable theoretical efforts \citep{Habouzit2022, Voit2024, Dattathri2025, PorrasValverde2026MassAssembly, Trinca2026}, the spin dimension has received comparatively less attention in models of galaxy formation, despite its far-reaching consequences for AGN feedback, gravitational wave astronomy, and the interpretation of multi-messenger observations. 

BH spin determines the efficiency with which accreted rest-mass energy is converted into radiation and jet power. For a geometrically thin, radiatively efficient disk, the radiative efficiency $\varepsilon_{\rm thin}$ ranges from $\approx 4\%$ for a maximally retrograde BH to $\approx 40\%$ for a maximally prograde one \citep{NovikovThorne1973}, compared to the fixed $\sim 10\%$ typically assumed in galaxy formation models. Separately, and especially in radiatively inefficient modes of accretion, the Blandford-Znajek (BZ) mechanism \citep{Blandford1977} extracts rotational energy from the BH via large-scale magnetic fields threading the horizon, powering relativistic jets with efficiencies that can exceed unity for near-maximal spins \citep{Tchekhovskoy2011, McKinney2012, Narayan2022}. The jet efficiency scales steeply with spin as $\eta_{\rm EM} \propto a_\star^2$, where $a_\star$ is the dimensionless BH spin parameter \citep{Blandford1977, ModerskiSikora1996, McKinney2004, Tchekhovskoy+2010, Cho2026}. Even modest changes in the spin distribution translate directly into order-of-magnitude changes in the radio jet power, and therefore, in the ability of AGN to quench star formation in massive galaxies.

At low Eddington ratios ($\dot{M} \lesssim 0.01$), the accreting gas cannot cool efficiently and inflates into a geometrically thick, radiatively inefficient flow known as an advection-dominated accretion flow \citep[ADAF;][]{NarayanYi1994, NarayanYi1995, Yuan&Narayan2014}. In this regime, large-scale magnetic flux can efficiently accumulate at the BH horizon. Once the accumulated flux reaches a saturated value $\phi$ and is strong enough to arrest the inflow, the disk enters the magnetically arrested disk \citep[MAD;][]{Narayan2003, Tchekhovskoy2011, Igumenshchev2003, Bisnovatyi-Kogan1974} state. A MAD is distinguished from the Standard and Normal Evolution \citep[SANE;][]{Narayan2012} state, in which the magnetic field remains relatively weak and disordered. In MAD states, the organized flux greatly amplifies the efficiency of the BZ jet relative to the SANE state. MAD BHs can launch jets with efficiencies well in excess of unity for near maximal spins \citep{Tchekhovskoy2011, McKinney2012, Narayan2022, Ricarte2023Spin}. We focus on the MAD regime because it represents the physical conditions under which BH spin has the strongest leverage on jet power, and therefore, on \textit{radio-mode} AGN feedback in massive galaxies. Semi-analytic models that fix the jet efficiency or assume a universal accretion mode miss this spin dependence entirely. {\sc Mad Sage} is designed to capture it self-consistently across cosmic time.

BH spin evolves through three main channels. Coherent disk accretion, where gas angular momentum is consistently aligned with the BH spin, drives spin-up and can bring a BH to near-maximal spin after accreting roughly its own mass \citep{Bardeen1970Nat, Thorne1974, ModerskiSikora1996, Gammie2004, Volonteri2005}. Incoherent (chaotic) accretion, where successive accretion episodes arrive with randomly oriented angular momenta, may drive spin-down towards low equilibrium values \citep{KingPringle2006, KingPringleHofmann2008, Fanidakis2011, Barausse2012}. BH-BH mergers contribute to both spin-up and spin-down depending on the progenitor spin and mass ratio, with the orbital angular momentum of the binary providing an additional spin-up contribution to the remnant \citep{Rezzolla2008, Berti2008, Hofmann2016, Varma2019}. Finally, in the MAD state, BZ jet launching itself drives spin-down by extracting rotational energy from the BH horizon, establishing an equilibrium spin that depends on the accretion rate and the disk state \citep{Tchekhovskoy+2010, Narayan2012, Skadowski2014, Narayan2022,Ricarte2023Spin}. At the ADAF equilibrium, this spin settles near $a_\star \approx 0.035$, while near-Eddington thin disk accretion settles near $a_\star \approx 0.7$-$0.8$.

Early semi-analytic models of BH spin evolution \citep{Volonteri2005, Berti2008, Barausse2012} showed that coherent accretion, chaotic accretion, and BH-BH mergers together produce a broad spin distribution across the galaxy population. These works show that massive BHs, which accrete more from the hot halo through randomly oriented Bondi inflow, tend to have lower spins than lower-mass BHs whose growth is dominated by coherent cold disk feeding. These models applied analytic innermost stable circular orbit (ISCO)-based efficiencies from the Novikov-Thorne thin-disk formula, not accounting for the role of magnetic fields in BH spin evolution and growth. 
They also did not couple spin evolution to the full galaxy formation context. More recently, \citet{Izquierdo-Villalba2020} extended {\sc L-Galaxies} semi-analytic model (SAM) with spin-dependent radiative efficiencies through the Novikov-Thorne ISCO formula \citep{NovikovThorne1973}, linking accretion coherence to host bulge kinematics through the $v/\sigma$ ratio. The model applies the thin-disk formula across all accretion rates, making no distinction between ADAF, thin-disk, or MAD regimes. No BZ jet torque is included. They find that BHs in elliptical galaxies have systematically lower spins than those in disk-dominated systems as a result of their more merger-dominated mass assembly histories. 

Recent hydrodynamical simulations have incorporated physically motivated spin-dependent feedback. \citet{Beckmann2025} tracked spin in {\sc NewHorizon} using resolved gas angular momentum to set the accretion direction, and adopted a \citet{McKinney2012} thick-disk solution in the low-Eddington regime with a spin-dependent BZ jet efficiency. Similar approaches coupling resolved angular momentum to BH spin evolution and jet feedback have been implemented in idealised circumnuclear disk settings \citep{Fiacconi2018, Talbot2021, Talbot2022}. \citet{Beckmann2025} find little evidence for efficient spin-down through chaotic accretion in a cosmological context. \citet{Husko2026} presented the {\sc COLIBRE} hybrid AGN feedback model, adopting the MAD assumption across three accretion disk states with spin-dependent BZ jet efficiencies from \citet{Tchekhovskoy+2010} and spin evolution following \citet{Narayan2022} and \citet{Ricarte2023Spin}. They find near-maximal spins for intermediate-mass BHs ($M_{\rm BH} \in [10^6, 10^8]\,\mathrm{M}_\odot$), declining at higher masses.

{\sc Mad Sage} extends previous semi-analytic treatments of BH spin by incorporating GRMHD-calibrated spin evolution and jet efficiencies into a galaxy formation framework. In contrast to models that describe spin evolution primarily through thin-disk accretion and BH mergers, we account for the competition between accretion-driven spin-up and Blandford-Znajek spin-down across different accretion regimes, following the prescriptions of \citet{Ricarte2023Spin}. We retain the Novikov-Thorne radiative efficiency while explicitly coupling the electromagnetic jet power to the evolving BH spin and accretion rate. Unlike \citet{Izquierdo-Villalba2020}, who relate accretion coherence to host bulge kinematics,
we associate coherent accretion with gravitational instabilities in the cold galactic disk and incoherent accretion with hot-halo Bondi feeding. The resulting accretion histories determine the evolution of BH spin, which in turn regulates the efficiency of AGN feedback and the subsequent evolution of the host galaxy. Our approach complements the hydrodynamical simulations of \citet{Beckmann2025} and \citet{Husko2026}, which resolve local gas dynamics but at a substantially greater computational cost. {\sc Mad Sage} instead allows efficient exploration of how galaxy-scale gas supply, BH spin evolution, and spin-dependent AGN feedback shapes galaxy populations across cosmic time.

The MAD prescription was placed on a firm theoretical footing by a series of General Relativistic Radiation Magnetohydrodynamic (GRRMHD) simulations that calibrated the BZ jet efficiency, radiative efficiency, and spin-down rate as functions of spin and accretion rate across the thick disk (ADAF), thin disk, and super-Eddington (slim disk) regimes \citep{Tchekhovskoy+2010, Tchekhovskoy2011, Narayan2022, Ricarte2023Spin}. \citet{Ricarte2023Spin} synthesized these results into a unified set of fitting functions applicable across all accretion regimes. These were implemented in the {\sc Serotina} SAM \citep{Ricarte+2025}, showing that the combined effect of MAD spin-down and coherent accretion produces a rich variety of spin signatures in the SMBH population. However, {\sc Serotina} focuses on the SMBH population in isolation from the galaxy evolution framework and the physics associated with BH mass assembly. This leaves open the question of how realistic galaxy-scale accretion histories alter the predicted spin distributions and their consequence for AGN feedback. 

In this paper, we present {\sc Mad Sage}, a new semi-analytic model of galaxy formation that self-consistently couples BH spin evolution to AGN feedback within the {\sc Dark Sage} galaxy formation framework \citep{Stevens2016, Stevens2024, Porras-Valverde2024, PorrasValverde2026MassAssembly}. 

This paper is organized as follows. Section~\ref{S:methods} describes the {\sc Mad Sage} SAM framework, the implementation of BH spin evolution, and the four model variants explored. Section~\ref{S:results} presents our main results. We first examine how spin evolution affects galaxy-scale observables, including the stellar mass function, BH scaling relations, AGN energy output, the AGN bolometric luminosity, and disk surface density profiles (Section~\ref{sec:galaxy_results}). We then present the predicted BH spin distributions across mass and cosmic time (Section~\ref{sec:spin_results}), and translate these into observational signatures for X-ray reflection spectroscopy, ngEHT/BHEX, and LISA (Section~\ref{sec:BHspin_observable}). Finally, we isolate the effect of BH-BH merger spin (Section~\ref{sec:BH-BHmergerspin}). Section~\ref{S:discussion} discusses the validity of the MAD assumption, the implications for jet launching mechanisms, and the radio-loud/radio-quiet dichotomy. We also place {\sc Mad Sage} in the broader landscape of spin models and compare with complementary observational probes of BH spin. We summarize our conclusions in Section~\ref{S:conclusions}. 

\section{Overview of {\sc Mad Sage}}
\label{S:methods}

\begin{figure*}
    \centering
    \includegraphics[width=\linewidth]{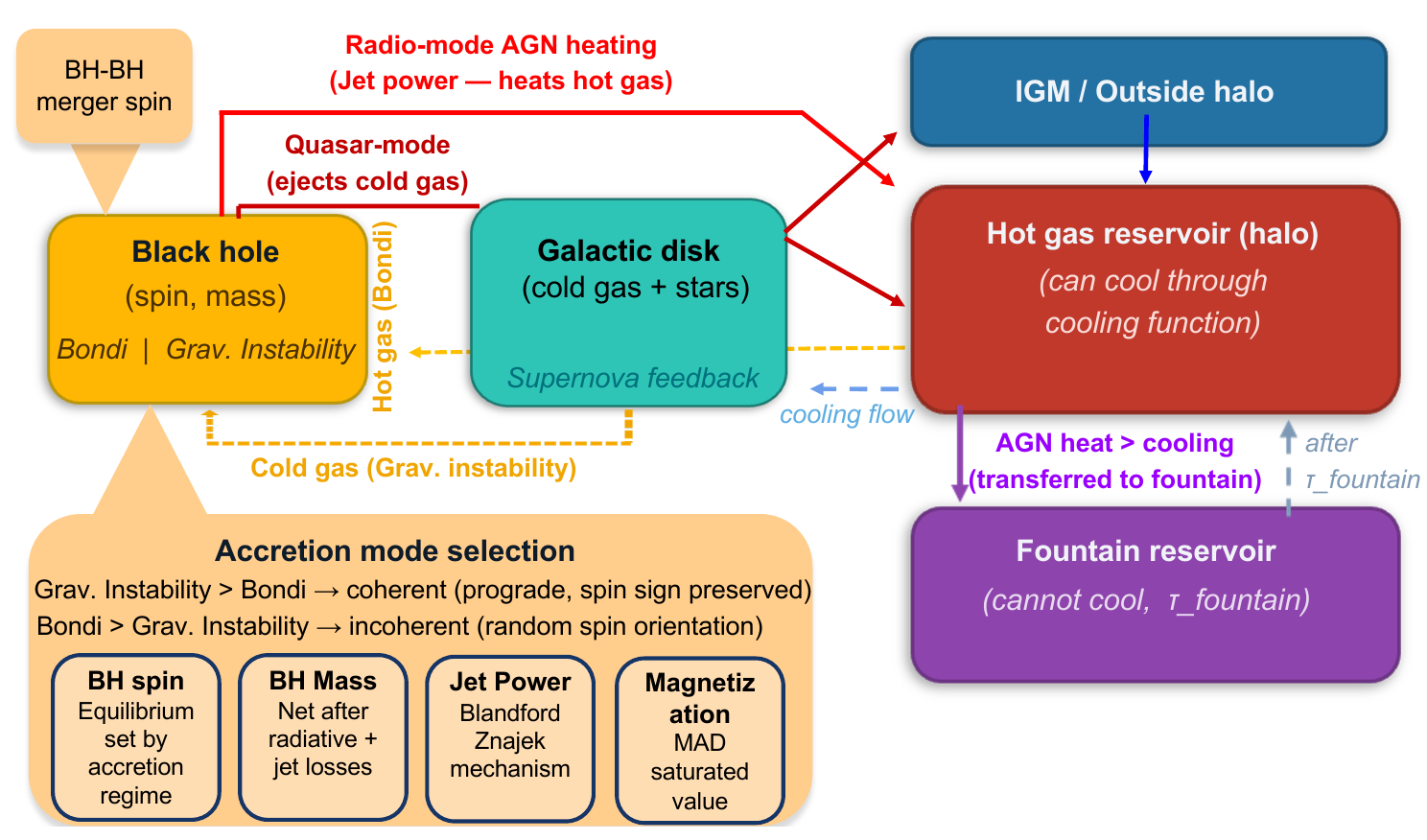}
    \caption{Schematic of the {\sc Mad Sage} semi-analytic model structure showing the black hole growth channels, spin and mass evolution, AGN feedback modes, and gas reservoirs that govern galaxy formation within the model.}
    \label{fig:MADSAGE_diagram}
\end{figure*}

We introduce the Magnetically Arrested Disk inclusion in the semi-analytic Galaxy Evolution model ({\sc Mad Sage}), built on the framework of {\sc Dark Sage} \citep{Stevens2016, Stevens2024} and its predecessor \citep{Croton2016}, with fundamental enhancements to black hole physics and AGN feedback. The key innovations are: (a) a decoupling of accretion channels from specific feedback modes: all accreted black hole mass now drives both {\it radio-mode} and {\it quasar-mode} feedback based on a unified and self-consistent AGN prescription; (b) self-consistent tracking of black hole spin evolution, accretion disk magnetization, and jet power from event-horizon to galactic scales; and (c) AGN feedback efficiency now evolves self-consistently with spin rather than being a fixed free parameter. These changes substantially alter how black holes and their host galaxies co-evolve. We provide a schematic illustration of these processes in Figure \ref{fig:MADSAGE_diagram}.

Following the standard semi-analytic approach, galaxy formation begins with hot gas that cools over a characteristic timescale \citep{White1978} and collapses into a gravitationally-supported disk \citep{Peebles1969, Fall1980, Fall1983, Kauffmann1996, MoMaoWhite1998, Bullock2001}. Within the disk, galaxies self-regulate their star formation through the balance between in-situ gas (originating within the galaxy) and ex-situ gas (accreted during galaxy mergers) \citep{DeLuciaBlaizot2007, Dekel2009Nat, Oser2010}. Supernova feedback from starbursts removes remaining gas unable to form stars, providing a critical constraint on the star formation efficiency \citep{Silk2001, Monaco2004, Bertone2007, Dubois2008}. Additionally, black holes co-evolve with their host galaxies. AGN feedback from accreting black holes further regulates cold gas availability and quenches star formation, particularly in massive systems \citep{SilkRees1998, Ferrarese2000, Kauffmann2000, King2003, Volonteri2003, DiMatteo2005Nat, Croton2006, DiMatteo2012}. Inhereted from {\sc Dark Sage} \citep{Stevens2016, Stevens2024}, {\sc Mad Sage} uses two annular disk structures (one stellar, one gaseous) to calculate star formation rates, disk instabilities, AGN and supernova feedback, and metallicity evolution.

The key physical advancement is implementing SMBH prescriptions from GRMHD simulations of \citet{Ricarte2023Spin}. Rather than using simplified spin models, we couple SMBH spin evolution directly to the accretion rate history. Specifically, we evaluate the magnetic flux parameter $\phi(a_\star, f_{\rm Edd})$, electromagnetic jet efficiency $\eta_{\rm EM}(a_\star, \phi)$, and spin-up rate $s(a_\star, f_{\rm Edd})$ from \citet{Ricarte2023Spin}, which derive from 38 three-dimensional GRRMHD simulations across sub- to super-Eddington accretion rates. This enables self-consistent spin evolution across accretion regimes: thin disks ($f_{\rm Edd} \lesssim 0.3$), super-Eddington flows ($f_{\rm Edd} \gtrsim 1$), and hot accretion ($f_{\rm Edd} \ll 0.01$). We additionally incorporate black hole merger spin evolution using analytic fits from \citet{Rezzolla2008}, capturing how coalescence events modulate spin distributions over cosmic time.

We run {\sc Mad Sage} on dark matter only merger trees from {\sc IllustrisTNG100-1} \citep{Pillepich2018, Springel2018FirstClustering, Nelson2018, Naiman2018, Nelson2019}, a periodic box of length $110.7$~cMpc with particle mass resolution $8.9 \times 10^6\, \mathrm{M}_{\odot}$ and Planck cosmology \citep{Planck2016}: $\Omega_M = 0.3089$, $\Omega_{\Lambda} = 0.6911$, $\Omega_b = 0.0486$, $h = 0.6774$.

\subsection{Black Hole Mass and Spin Evolution}\label{BHmass_spinevol}

{\sc Mad Sage} seeds $100\, \mathrm{M}_{\odot}$ BHs in every $\geq 10^9\, \mathrm{M}_{\odot}$ halo in the underlying N-body simulation. Black holes grow through two channels. In the first, hot halo gas accretes following a Bondi-Hoyle prescription \citep{Bondi1952} combined with the `maximal cooling flow' model of \citet{Nulsen2000MNRAS}. The hot-gas accretion rate is

\begin{equation}
\label{eq:bondi}
\dot{M}_{\rm BH, \ Bondi} = \frac{15 \uppi \, G \, \bar{\mu} m_{\rm p}\, k_{\rm B} \, T_{\rm vir}}{16\, \Lambda(T_{\rm vir}, Z_{\rm hot})} M_{\rm BH}
\end{equation}

\noindent where $\bar{\mu}\, m_p\,$ is the mean particle mass ($\bar{\mu} = 0.59$ for a fully ionized gas), $k_{\rm B}$ is the Boltzmann constant, $T_{\rm vir} = 35.9\,V_{\rm vir}^2\,{\rm K}$ is the virial temperature of the hot halo, and $\Lambda (T_{\rm vir},Z_{\rm hot})$ is the metal-dependent cooling function. This channel is treated as incoherent, as hot halo gas is assumed to arrive with random angular momentum orientation. For this channel, the accretion rate is capped at the Eddington limit.

The second channel is gravitational instability accretion from the cold galactic disk, treated as coherent because disk gas is assumed to carry a fixed prograde angular momentum. Cold gas accretes when a disk annulus becomes Toomre-unstable \citep{Toomre1964}. For each annulus $i$ spanning radii $r_i$ to $r_{i+1}$, the Toomre parameter for the gas is

$$Q_{\rm gas}^i = \frac{c_s\,\kappa_i\,(r_{i+1}^2 - r_i^2)}{G\,M_{\rm gas}^i}$$

\noindent where $\kappa_i$ is the epicyclic frequency, $c_s$ is the cold-gas sound speed, and $M_{\rm gas}^i$ is the gas mass in the annulus. When stars and gas coexist, their combined stability is assessed through the two-component criterion \citep{Romeo2011}:

$$Q_{\rm tot}^{-1} = W\,Q_{\rm gas}^{-1} + Q_\star^{-1}, \qquad W = \frac{2\sigma_R c_s}{\sigma_R^2 + c_s^2}$$

\noindent where $\sigma_R$ is the radial stellar velocity dispersion. When $Q_{\rm tot} < Q_{\rm min}$, the unstable gas mass in annulus $i$ is

$$\Delta M_{\rm unstable}^i = M_{\rm gas}^i\left(1 - \frac{Q_{\rm gas}^i}{Q_{\rm min}}\right)$$

\noindent where $Q_{\rm min} = 1$ is the critical stability threshold. A fraction $\kappa_{\rm sink}$ (gas sink rate) of $\Delta M_{\rm unstable}$ flows inward through angular momentum transport, conserving angular momentum across adjacent annuli. Gas that cascades to the innermost annulus is deposited into the BH coherent reservoir. We note that the coherent/incoherent classification is orthogonal to the Eddington-ratio regime. A BH accreting incoherently from hot halo gas may still reach $f_c < f_{\rm Edd} < 1$, and coherent disk-fed accretion may fall below $f_c$. The two classifications operate independently. 

For each accretion channel and timestep, given a gas inflow rate $\dot{M}$, we compute the instantaneous Eddington ratio $f_{\rm Edd} = \dot{M}/\dot{M}_{\rm Edd}$ and track the dimensionless spin $a_\star \in [-0.998,\,0.998]$, where the bounds correspond to the maximum allowed spin magnitude due to radiative capture effects during thin-disk accretion \citep{Thorne1974}. The MAD BH recipes hereafter have also been implemented in the {\sc Serotina} SAM \citep{Ricarte+2025}, but in {\sc Mad Sage}, $\dot{M}$ is calculated based on galaxy-level astrophysics instead of fully prescribed. In addition, while {\sc Serotina} lacks gas entirely, the calculated feedback is now coupled self-consistently with the galaxy in {\sc Mad Sage}.

The spin evolves according to \citep{Gammie2004, Shapiro2005}

\begin{equation}
    \frac{da_\star}{dt} = s\,\frac{\dot{M_{\rm }}}{M},
    \label{eq:spin_evolution}
\end{equation}

\noindent where $s$ is the dimensionless spin-up parameter, defined generally as $s = l - 2a_\star e$, with $l$ the inward specific angular momentum flux and $e$ the inward specific energy flux of accreted gas, both evaluated at $5\,r_{\rm g}$ \citep{Ricarte2023Spin}. Following equation~(13) of \citet{Ricarte2023Spin}, $s$ is prescribed as

\begin{equation}
s = \begin{cases}
s_{\rm MAD} & f_{\rm Edd} \leq f_{\rm c}, \\[4pt]
s_{\rm HD} + s_{\rm EM} & f_{\rm c} < f_{\rm Edd} \leq 1, \\[4pt]
s_{\rm SE} & f_{\rm Edd} > 1,
\end{cases}
\label{eq:sigma_total}
\end{equation}

\noindent where $f_{\rm c} \equiv 0.01$ is the hot-accretion threshold. The hydrodynamic spin-up component is

\begin{equation}
    s_{\rm HD} = l_{\rm HD} - 2a_\star e_{\rm HD},
    \label{eq:sHD}
\end{equation}

\noindent representing angular momentum deposited by the infalling gas at the ISCO, and the electromagnetic spin-down component is

\begin{equation}
    s_{\rm EM} = \mathrm{sign}(a_\star)\,\eta_{\rm EM}
    \!\left(\frac{1}{k\,\Omega_{\rm H}} - 2a_\star\right),
    \label{eq:sEM}
\end{equation}

\noindent representing angular momentum extracted by the Blandford-Znajek jet, where $\eta_{\rm EM}$ is the electromagnetic jet efficiency, $\Omega_{\rm H} = |a_\star|/[2(1+\sqrt{1-a_\star^2})]$ is the angular velocity of the BH horizon, and $k$ is the ratio of the angular frequency of the field lines to that of the BH. Following \citet{Lowell2024} and \citet{Ricarte2023Spin}, we adopt

\begin{equation}
k(a_\star) = \begin{cases}
0.23, & a_\star < 0,\\
\min(0.1 + 0.5\,a_\star,\;0.35), & a_\star \geq 0.
\end{cases}
\label{eq:k}
\end{equation}

\noindent A fitting function for $s_{\rm HD}$ interpolates smoothly between the Novikov-Thorne thin-disk solution as $f_{\rm Edd} \to 0$ and nonradiative GRMHD result as $f_{\rm Edd} \to \infty$ \citep{Ricarte2023Spin}.

Below the threshold ($f_{\rm Edd} \leq f_{\rm c}$), the flow becomes a hot, geometrically thick ADAF and $s_{\rm MAD}$ is replaced by a fitting function to ideal GRRMHD simulations \citep{Narayan2022}, 

\begin{align}
    s_{\rm MAD}(a_\star) = \; & 0.45 - 12.53\,a_\star - 7.80\,a_\star^2 \nonumber \\
    & + 9.44\,a_\star^3 + 5.71\,a_\star^4 - 4.03\,a_\star^5.
    \label{eq:sMAD}
\end{align}

\noindent which drives the spin toward an equilibrium $a_\star \approx 0.035$. Above the Eddington limit ($f_{\rm Edd} > 1$), the disk again becomes geometrically thick. $s_{\rm SE}$ is replaced by a fitting function to nonradiative GRMHD simulations \citep{Ricarte2023Spin} which drives the spin towards an equilibrium value of $a_\star \approx 0.035$. This approach naturally transitions between regimes. Thin disks drive rapid spin-up toward $a_\star \approx 0.998$, and near-Eddington systems undergo mild spin-down to equilibrium value of $a_\star \approx 0.7$.

The electromagnetic jet power is computed via the Blandford-Znajek mechanism \citep{Blandford1977},

\begin{equation}
    P_{\rm jet} = \eta_{\rm EM}(a_\star,\,\phi)\,\dot{M}c^{2},
    \label{eq:jet_power}
\end{equation}

\noindent where the jet efficiency is

\begin{equation}
    \eta_{\rm EM} = \frac{\kappa}{4\pi}\,\phi^{2}\,\Omega_{\rm H}^{2}
        \left[1 + 1.38\,\Omega_{\rm H}^{2} - 9.2\,\Omega_{\rm H}^{4}\right],
    \label{eq:eta_EM}
\end{equation}

\noindent with $\kappa = 0.05$, and $\Omega_{\rm H}$ is again the angular velocity of the event horizon \citep{Ricarte2023Spin}. The dimensionless magnetic flux $\phi$ threading the horizon is set by the MAD saturation value modulated by the Eddington ratio,

\begin{equation}
    \phi(a_\star, f_{\rm Edd}) =
        \phi_{\rm MAD}(a_\star)\,
        \frac{(f_{\rm Edd}/f_{\rm crit})^{\alpha}}
             {1 + (f_{\rm Edd}/f_{\rm crit})^{\alpha}},
    \label{eq:phi}
\end{equation}

\noindent where $\alpha = 1.29$ is a fitting parameter calibrated to GRMHD simulations of the MAD-thin disk transition, $f_{\rm crit} = 1.88$ is the critical Eddington ratio at which $\phi$ reaches half its maximum value, and

\begin{equation}
    \phi_{\rm MAD}(a_\star) = 52.6 + 34\,a_\star
        - 14.9\,a_\star^{2} - 20.2\,a_\star^{3}
    \label{eq:phi_MAD}
\end{equation}

\noindent is the spin-dependent saturation flux from nonradiative MAD simulations \citep{Ricarte2023Spin}. By construction, $\phi \to \phi_{\rm MAD}/2$ as $f_{\rm Edd} \to 0$ from above (approaching the thin-disk regime) and $\phi \to \phi_{\rm MAD}(a_\star)$ as $f_{\rm Edd} \to \infty$ (nonradiative thick disk limit). 

With the spin-up parameter and magnetic flux fully specified as functions of $a_\star$ and $f_{\rm Edd}$, equation~(\ref{eq:spin_evolution}) is solved self-consistently for each accretion event. The spin-up parameter $s$ applies in both accretion modes. The distinction between modes lies in how the spin direction is handled. In the coherent mode (cold gas supply exceeds hot-gas supply), the spin sign is preserved throughout the integration, consistent with the assumption that disk gas carries a fixed prograde angular momentum. In incoherent mode, the spin magnitude $|a_\star|$ is evolved using the same $s$, but after each sub-step, the spin sign is randomly reassigned to prograde or retrograde with equal probability, accounting for the unknown angular momentum orientation of the infalling hot gas. Both accretion channels feed a single unified spin and mass evolution step, where an adaptive Runge-Kutta-Fehlberg 4(5) integrator simultaneously solves equations~(\ref{eq:spin_evolution})-(\ref{eq:phi}) over the sub-step timestep $\Delta t$. We checked that this physically reassignment of spin does not depend on the adaptive integration step size.

The Eddington ratio is re-evaluated at each Runge-Kutta stage as the black hole mass grows. The integrator returns the updated spin $a_\star$ and net accreted mass $\Delta M$ at the end of the timestep, together with the time-averaged jet power $P_{\rm jet}$, and the disk magnetisation $\phi$. These outputs drive the AGN feedback from the event-horizon scales to galactic scales. 

\subsection{AGN feedback}

AGN feedback is applied in two modes: {\it radio-mode} which is preventative and acts on the hot gas reservoir and {\it quasar-mode}, which ejects cold gas from the galactic disk. In \textit{radio-mode}, the jet power computed from the Blandford-Znajek mechanism can be very small for black holes with low spin, since $\eta_{\rm EM} \propto \Omega_{\rm H}^{2} \propto a_\star^{2}$ vanishes as $a_\star \to 0$. Because radiatively inefficient accretion flows can also transfer energy to the surrounding gas through winds, convection, and magnetic stresses, even at low BH spin \cite{Yuan&Narayan2014,Cho+2023,Cho+2024}, we introduce a phenomenological floor on the effective mechanical power available for halo heating:

\begin{equation}
    \dot{E}_{\rm mech} = \max\!\left(P_{\rm jet},\;
                   \eta_{\rm floor}\,\dot{M}c^{2}\right),
    \label{eq:jet_energyetafloor}
\end{equation}

\noindent where $\dot{E}_{\rm mech}$ is the effective mechanical power coupled to the halo which includes additional energy beyond BZ. We assume that $\eta_{\rm floor} = 0.02$, motivated by GRMHD simulations of ADAF flows around non-spinning BHs that find minimum outflow efficiency of $\eta \approx 0.02$ \citep{Cho+2023, Cho+2024}, driven by disk winds, convection, and magnetic pressure, independent of the Blandford-Znajek mechanism. The jet energy heats the hot gas reservoir, suppressing cooling. When jet heating exceeds the instantaneous cooling rate, the excess hot gas is transferred to a fountain reservoir, where it cannot cool for a dynamical timescale $\tau_{\rm fountain}$, preventing runaway star formation in massive galaxies. This fountain reservoir is a feature unique to {\sc Dark Sage} among semi-analytic models \citep{Stevens2024}. It replaces the $r_{\rm heat}$ heating-radius mechanism of \citet{Croton2006, Croton2016}, which suppresses cooling within a fixed spatial radius. Instead, recooling is delayed by a timescale tied to the halo dynamical time. In {\it quasar-mode}, the energy

\begin{equation}
   \dot{E}_{Q} = \varepsilon_{\rm thin}(a_{\star})\,\dot{M}c^{2}
\end{equation}

\noindent ejects cold gas from the galactic disk to the hot halo or beyond to the circumgalactic medium, where $\varepsilon_{\rm thin}(a_{\star})$ is the Novikov-Thorne radiative efficiency evaluated at the current spin \citep{NovikovThorne1973}.

\begin{table*}
\centering
\setlength{\tabcolsep}{4pt}
\begin{tabular}{lllcccc}
\toprule
Model &
\textit{Radio-mode} &
\textit{Quasar-mode} &
$\eta_{\rm floor}$ &
BH-BH merger spin &
$\kappa_{\rm sink}$ &
$E_{\rm SN}$ [erg] \\
\midrule
\textbf{No-spin} &
$\eta = 0.3$ (fixed) &
$\eta = 0.3$ (fixed) &
No &
- &
$0.70$ &
$10^{50}$ \\
\midrule
\textbf{Coherent only} &
$\eta_{\rm EM}(a_\star, \phi)$ &
$\varepsilon_{\rm thin}(a_\star)$ &
No &
No &
$0.70$ &
$7.03 \times 10^{50}$ \\
\midrule
\textbf{Full-physics} &
$\eta_{\rm EM}(a_\star, \phi)$ &
$\varepsilon_{\rm thin}(a_\star)$ &
No &
Yes &
$0.85$ &
$1.31 \times 10^{50}$ \\
\midrule
\textbf{Full-physics + correction} &
$\eta_{\rm EM}(a_\star, \phi)$ &
$\varepsilon_{\rm thin}(a_\star)$ &
$0.02$ &
Yes &
$0.82$ &
$8.26 \times 10^{50}$ \\
\bottomrule
\end{tabular}
\caption{Summary of the models explored in this work. In the \textbf{No-spin} model, Bondi accretion is coupled exclusively to \textit{radio-mode} feedback and gravitational instability accretion is coupled exclusively to \textit{quasar-mode} feedback, with fixed radiative efficiencies. In all models that track BH spin (i.e., all models except \textbf{No-spin}), both BH accretion channels contribute to \textit{radio-mode} and \textit{quasar-mode} feedback, removing this one-to-one coupling. The \textit{radio-mode} efficiency is set by the self-consistently evolved electromagnetic jet efficiency $\eta_{\rm EM}(a_\star, \phi)$ via the Blandford-Znajek mechanism \citep[equation~\ref{eq:eta_EM};][]{Ricarte2023Spin}, and the \textit{quasar-mode} efficiency by the Novikov-Thorne thin-disk radiative efficiency $\varepsilon_{\rm thin}(a_\star)$ \citep{NovikovThorne1973}, both of which vary with the instantaneous spin $a_\star$. All spin models include mass losses from radiative and jet processes during accretion, which are absent in the \textbf{No-spin} model. The \textbf{Full-physics + correction} model imposes a minimum effective mechanical feedback efficiency $\eta_{\rm floor}$ (equation~\ref{eq:jet_energyetafloor}) accounting phenomenologically for nonjet outflows from radiatively inefficient accretion flows. Where included, BH-BH merger spin is computed using the \citet{Rezzolla2008} fitting formulae with isotropically sampled spin orientations (Section~\ref{sec:BHmerger_spin}). The free parameters $\kappa_{\rm sink}$ (gas sink rate) and $E_{\rm SN}$ (supernova energy) are calibrated independently and optimized against the observed stellar mass function at $z = 0$.}
\label{tab:tablemodels}
\end{table*}

\subsection{ {\rm BH-BH} merger spin}\label{sec:BHmerger_spin}

When two galaxies merge after their host subhalos have coalesced on the dynamical friction timescale \citep{BinneyTremaine1987}, their central black holes are assumed to merge instantaneously. The spin of the remnant black hole is computed using the analytic fitting formulae of \citet{Rezzolla2008}, which were calibrated against numerical relativity simulations of binary black hole mergers spanning a range of mass ratios and spin configurations. The remnant spin depends on the mass ratio $q = M_{2}/M_{1} \leq 1$, the spin magnitudes $a_{*,1}$ and $a_{*,2}$ of the two progenitors, and the orientations of their spin vectors relative to the orbital angular momentum of the binary. The total angular momentum of the system is

\begin{equation}
    \mathbf{J}_{\rm total} = \mathbf{L}_{\rm orb}
                            + \mathbf{S}_{1} + \mathbf{S}_{2},
    \label{eq:J_total}
\end{equation}

\noindent where $\mathbf{L}_{\rm orb}$ is the orbital angular momentum and
$\mathbf{S}_{i} = a_{*,i} M_{i}^{2}$ are the spin angular momenta of
each progenitor. The direction of $\mathbf{J}_{\rm total}$ sets the spin axis of the remnant, and its magnitude determines the final spin $a_{*,\rm rem}$. The sign of the remnant spin can therefore differ from that of either progenitor depending on the relative orientation of the three angular
momentum vectors.

Because the orientation of each progenitor spin relative to the orbital plane is unconstrained at the resolution of a semi-analytic model, we draw the polar angles $\theta_{1}$ and $\theta_{2}$, which measure the tilt of each spin axis with respect to $\mathbf{L}_{\rm orb}$, from an isotropic distribution,

\begin{equation}
    \theta_{i} = \arccos(1 - 2\mathcal{U}_{i})\times
                 \mathrm{sign}(\mathcal{U}_{i}^{\prime} - 0.5),
    \label{eq:theta_sampling}
\end{equation}

\noindent where $\mathcal{U}_{i}$ and $\mathcal{U}_{i}^{\prime}$ are independent draws from a uniform distribution on $[0,1]$. The azimuthal angles $\varphi_{1}$ and $\varphi_{2}$, which describe the precession of each spin around $\mathbf{L}_{\rm orb}$, are drawn uniformly from $[0, 2\pi)$. Together, the four angles $(\theta_{1}, \varphi_{1}, \theta_{2}, \varphi_{2})$ sample all possible spin orientations uniformly, implementing an isotropic prior on the angular momentum geometry of the binary. The remnant mass is set to $M_{\rm rem} = M_{1} + M_{2}$, neglecting the small fraction of mass radiated as gravitational waves.

\subsection{Model Variations}
\label{sec:model_variations}

Table~\ref{tab:tablemodels} summarizes the four models explored in this work, which are designed to isolate the physical effects of spin evolution, accretion mode, and BH-BH merger spin on galaxy and black hole co-evolution.

The \textbf{No-spin} model follows a refactored version of the standard {\sc Dark Sage} prescription \citep{Stevens2016, Stevens2024}. In the original {\sc Dark Sage} prescription, Bondi accretion and gravitational instability accretion were coupled one-to-one to \textit{radio-mode} and \textit{quasar-mode} feedback respectively. Both channels adopted a fixed radiative efficiency of $\eta = 0.3$. This implicitly assumed that BHs switch instantaneously between accretion modes within a sub-time step. No spin evolution is tracked, and BH-BH merger spin is not modelled. This model serves as our reference against which all spin model predictions are compared. In {\sc Mad Sage}, we decouple these channels. Both Bondi and gravitational instability accretion contribute to a single unified BH mass and spin evolution step. The resulting jet power and radiative efficiency then drive both feedback modes self-consistently, based on the instantaneous spin and accretion state of the black hole.

The \textbf{Coherent only} model introduces self-consistent BH spin evolution but assumes that all accretion proceeds in the coherent regime, regardless of whether the accreting gas originates from the hot halo or the cold disk. This is equivalent to assuming that all infalling gas carries a fixed prograde angular momentum aligned with the BH spin axis. Both {\it radio-mode} and {\it quasar-mode} feedback efficiencies evolve with spin via $\eta_{\rm EM}(a_\star, \phi)$ and $\varepsilon_{\rm thin}(a_\star)$ respectively. No $\eta_{\rm floor}$ is applied and BH-BH merger spin is not included, allowing us to isolate the effect of accretion-driven spin evolution alone.

The \textbf{Full-physics} model relaxes the assumption of purely coherent accretion. As stated in section~\ref{BHmass_spinevol}, hot gas accretion via the Bondi channel is treated as incoherent, with the spin sign randomly reassigned after each integration step to reflect the isotropic angular momentum of the hot halo. Cold disk accretion via gravitational instability remains coherent. The dominant channel at each timestep determines the accretion mode. When the cold gas supply exceeds the hot gas supply, accretion proceeds in coherent mode. Otherwise, incoherent mode is used (Section~\ref{S:methods}). BH-BH merger spin is included in this model using the \citet{Rezzolla2008} formulae, so that both accretion-driven and merger-driven spin evolution are tracked self-consistently. No $\eta_{\rm floor}$ is applied. 

The \textbf{Full-physics + correction} model, is identical to the \textbf{Full-physics} model but additionally imposes the minimum jet energy floor $\eta_{\rm floor} = 0.02$ (equation~\ref{eq:jet_energyetafloor}). 

\subsection{Calibration}

Each model is calibrated using two steps. First, we run {\sc Mad Sage} on the low-resolution {\sc TNG50-4} merger trees and apply the swarm optimization algorithm described in \citet{Stevens2024} to identify the values of the free parameters (the gas sink rate $\kappa_{\rm sink}$ and the supernova energy $E_{\rm SN}$) that best reproduce the observed stellar mass function at $z=0$. Second, we rerun the model on the full {\sc TNG100-1} merger trees and further hand-calibrate the parameters to obtain the best fit to the observations shown in Figure~\ref{fig:SMF_z0}. The resulting parameter values for each model are listed in Table~\ref{tab:tablemodels}. We note that for the \textbf{Full-physics} model, no combination of free parameters could adequately reproduce the observed number density of massive galaxies at the high-mass end ($M_\star \gtrsim 10^{11}\,\mathrm{M}_\odot$), reflecting the insufficient AGN heating at low spin discussed in Section~\ref{sec:galaxy_results}. This motivated the introduction of the mechanical feedback floor $\eta_{\rm floor} = 0.02$ (informed by \cite{Cho+2023,Cho+2024}), which proved necessary to achieve a satisfactory fit to the high-mass end of the stellar mass function.

\section{Results}
\label{S:results}

Our implementation critically assumes that MAD conditions are achieved at the event horizon across all accretion rates and spin values. Under this assumption, the magnetic flux threading the BH is determined solely by the accretion rate and spin. We discuss the validity of this assumption in Section~\ref{S:discussion}.

\subsection{How does BH spin affect galaxy evolution?}
\label{sec:galaxy_results}

We now examine how the self-consistent coupling between BH spin evolution and AGN feedback shapes the predicted galaxy population, focusing on the stellar mass function, BH-host galaxy scaling relations, AGN energy output, AGN bolometric luminosity, and the radial structure of the gas and stellar disks.

\begin{figure}
    \centering
    \includegraphics[width=\linewidth]{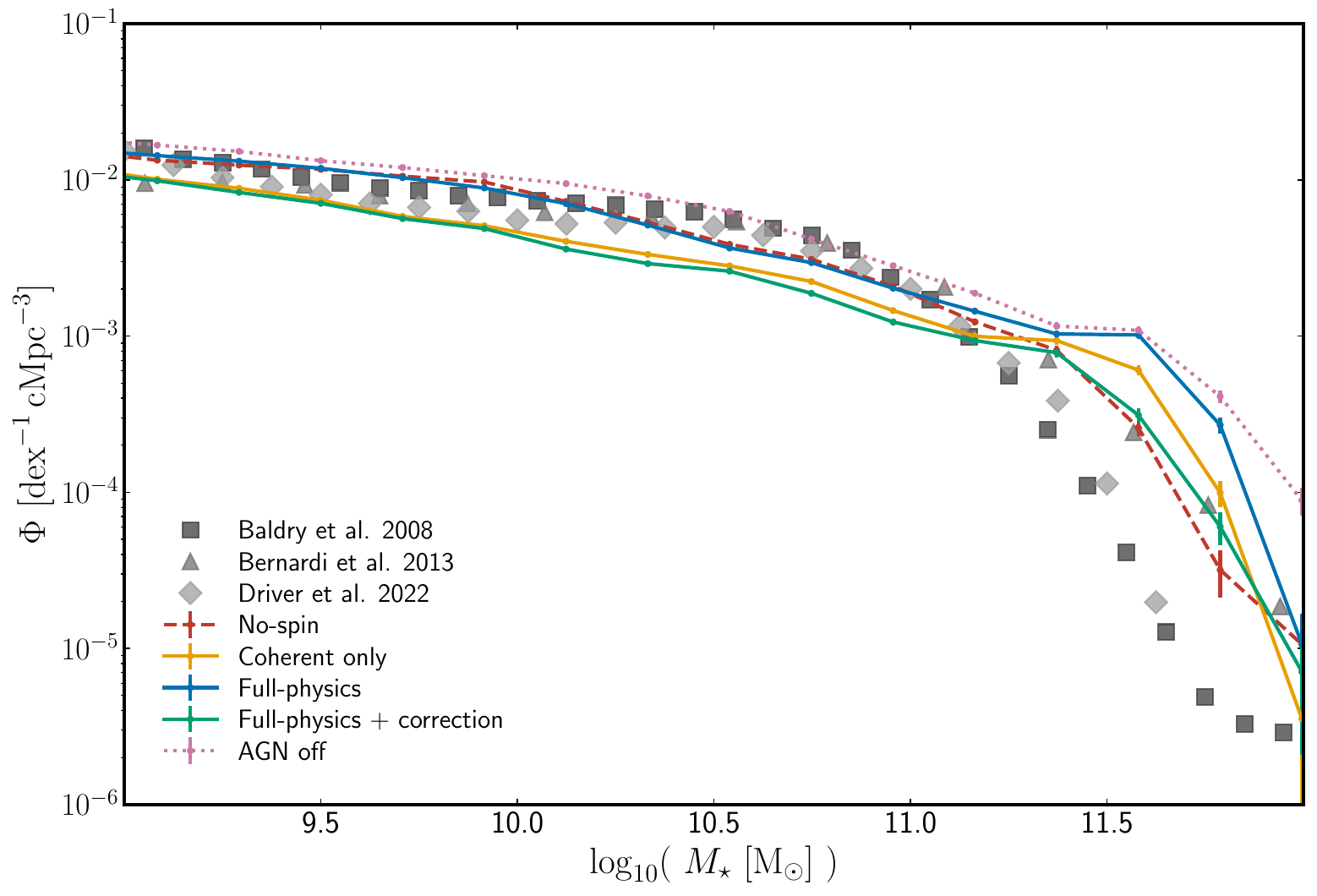}
    \caption{Stellar mass function at $z = 0$ for all models explored in this work. The dashed crimson line shows the \textbf{No-spin} model, the orange line shows the \textbf{Coherent only} model, the blue line shows the \textbf{Full-physics} model, the green line shows the \textbf{Full-physics + correction} model, and the dotted pink line shows a run with AGN feedback turned off. Observational estimates from \citet{Baldry2008}, \citet{Bernardi2013}, and \citet{Driver2022} are shown as grey squares, triangles, and diamonds, respectively. The \textbf{Full-physics} model overpredicts the number density of massive galaxies ($M_\star \gtrsim 10^{11}\,\mathrm{M}_\odot$) relative to the \textbf{No-spin} model, encoding weaker AGN quenching. This served as motivation to include the $\eta_{\rm floor}$ correction. The \textbf{Coherent only} and \textbf{Full-physics + correction} model are in reasonable agreement with observations.}
    \label{fig:SMF_z0}
\end{figure}

Figure~\ref{fig:SMF_z0} shows the stellar mass function at $z = 0$ for all our models. Our \textbf{No-spin} model provides a good match to the observed stellar mass function across the full mass range, consistent with its original calibration in \citet{Stevens2024}. As expected, turning off AGN feedback yields a higher number density of galaxies, particularly at the high-mass end, where the absence of any energy mechanism offsetting the cooling time allows hot gas to cool almost instantaneously and become available for star formation and black hole growth. 

Interestingly, the model closest to the AGN-off case is the \textbf{Full-physics} model. In this model, the incoherent Bondi channel dominates for black holes in massive galaxies at low redshift, randomly resetting the spin sign and driving BH spins toward the ADAF equilibrium value of $a_\star \approx 0.035$. At these low spins, the Blandford-Znajek efficiency $\eta_{\rm EM} \propto a_\star^2$ is orders of magnitude below the \textbf{No-spin} model value of $\eta = 0.3$ and is hence strongly suppressed relative to the fixed-efficiency baseline, revealing that only a small fraction of the accreted rest-mass energy is converted to jet power per accretion event \footnote{The lower radiative losses at low spin allow more mass to be retained by the BH, driving higher BH masses and consequently higher Bondi accretion rates ($\dot{M}_{\rm Bondi} \propto M_{\rm BH}^2$), which more than compensate for the low per-event efficiency and produce a higher absolute jet power compared to the \textbf{Coherent only} model at low redshift}. 

Conversely, the \textbf{Coherent only} model produces a better match to \citet{Bernardi2013} at the high-mass end. In this model, all accretion is treated as coherent regardless of channel, systematically spinning up BHs to high $|a_\star|$. The higher Novikov-Thorne radiative efficiency $\varepsilon_{\rm thin}(a_\star)$ at high spin raises the \textit{quasar-mode} energy output. However, per unit stellar mass, the \textit{quasar-mode} energy in the \textbf{Coherent only} model closely traces that of the \textbf{Full-physics + correction} model (see Appendix~\ref{app:AGNenergy_logTSM}). At the same time, the elevated radiative efficiency reduces the fraction of accreted mass retained by the BH, yielding lower average BH masses at fixed stellar mass and lower Bondi accretion rates. Note that differences in the free parameter $\kappa_{\rm sink}$ between models may also contribute to this offset. Despite reaching the highest BH spins of all models, the \textbf{Coherent only} model therefore produces the weakest \textit{radio-mode} energy at intermediate stellar masses. High spin raises the disk radiative losses and suppresses BH mass growth, reducing Bondi accretion rates that drive the jets. The gap narrows at the highest BH masses ($M_{\rm BH} \gtrsim 10^8\,\mathrm{M}_\odot$), where the BZ efficiency scaling $\eta_{\rm EM} \propto a_\star^2$ boosts jet power for the most massive, rapidly spinning BHs. The quenching differences at the high-mass end of the SMF therefore arise from a combination of the \textit{radio-mode} differences between spin models, the large scatter in BH mass at fixed stellar mass \citep[which can wash out the imprint of the feedback implementation
on the quenched mass function;][]{Porras-ValverdeForbes2025, Porras-ValverdeForbes2026}, and the upturn in \textit{quasar-mode} energy at high stellar masses in the \textbf{Coherent only} and \textbf{Full-physics + correction} models (see Appendix~\ref{app:AGNenergy_logTSM}).

\begin{figure*}
    \centering
    \includegraphics[width=\linewidth]{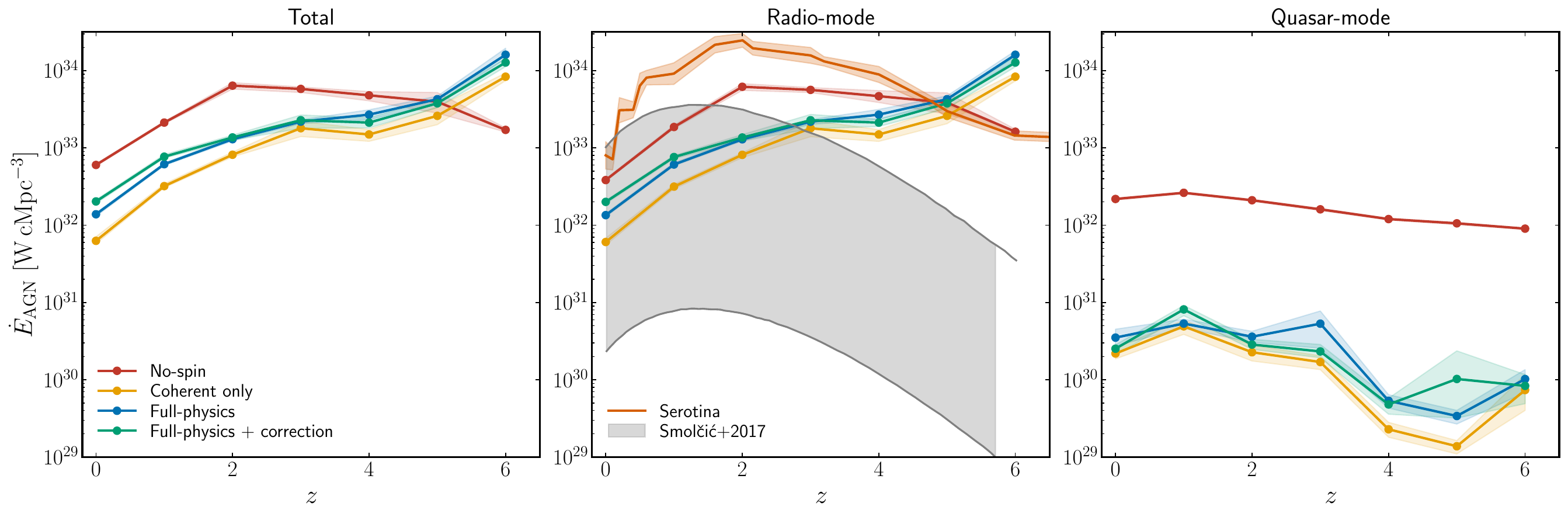}
    \caption{AGN energy density as a function of redshift for all models described in Table~\ref{tab:tablemodels}, decomposed into total energy output (left), jet power from {\it radio-mode} feedback (middle), and {\it quasar-mode} energy (right). Shaded bands show the 68th percentile bootstrap uncertainty from the simulation volume. The grey shaded region shows the observed kinetic energy density from radio-selected AGN compiled by \citet{Smolcic2017}, which represents an observational lower bound on the total jet power. The orange line shows the total kinetic density from the \textsc{Serotina} semi-analytic model \citep{Ricarte+2025}. All spin models predict lower total AGN energy densities than the \textbf{No-spin} model at $z \lesssim 5$. Among the spin models, the \textbf{Full-physics + correction} model produces slightly more energy output across all redshifts. This excess is driven by the self-consistent spin evolution combined with the minimum jet coupling enforced by $\eta_{\rm floor}$.}
    \label{fig:JetPower_redshift}
\end{figure*}

The \textbf{Full-physics} model illustrates an important distinction between the total AGN energy output and the effectiveness of AGN feedback in suppressing star formation. Despite its relatively high total AGN energy output (Figure~\ref{fig:JetPower_redshift}), this model substantially overproduces massive galaxies (Figure~\ref{fig:SMF_z0}). This behavior arises from the coupled evolution of BH spin, accretion rate, and feedback efficiency. Incoherent hot-halo accretion, together with Blandford-Znajek spin-down, drives massive BHs toward low spins, where the electromagnetic jet efficiency is strongly suppressed. At the same time, the reduced radiative efficiency allows BHs to retain a larger fraction of the accreted mass, increasing their subsequent Bondi accretion rates. The higher accretion rates can partially compensate for the low jet efficiency, producing substantial AGN energy output without providing sufficient heating to prevent cooling and star formation in massive halos.

This result highlights a limitation of representing \textit{radio-mode} AGN feedback solely through the electromagnetic power extracted by the Blandford-Znajek mechanism. Radiatively inefficient accretion flows can also drive winds and other mechanical outflows that transfer energy to the surrounding gas, even when the BH spin is small \cite{Narayan2012, Skadowski2013, Yuan&Narayan2014}. These contributions are not explicitly included in the Blandford-Znajek jet prescription. To account phenomenologically for these additional feedback channels, we included a term $\eta_{\rm floor}$ for mechanical feedback. This correction has a relatively modest effect on the integrated AGN energy budget but provides additional halo heating where the spin-dependent jet efficiency is lowest. As a result, the \textbf{Full-physics + correction} model produces stronger quenching of massive galaxies and better agreement with the observed high-mass end of the stellar mass function. Because the models are calibrated independently, the quantitative differences also reflect variations in the galaxy-formation parameters.

\begin{figure}
    \centering
    \includegraphics[width=\linewidth]{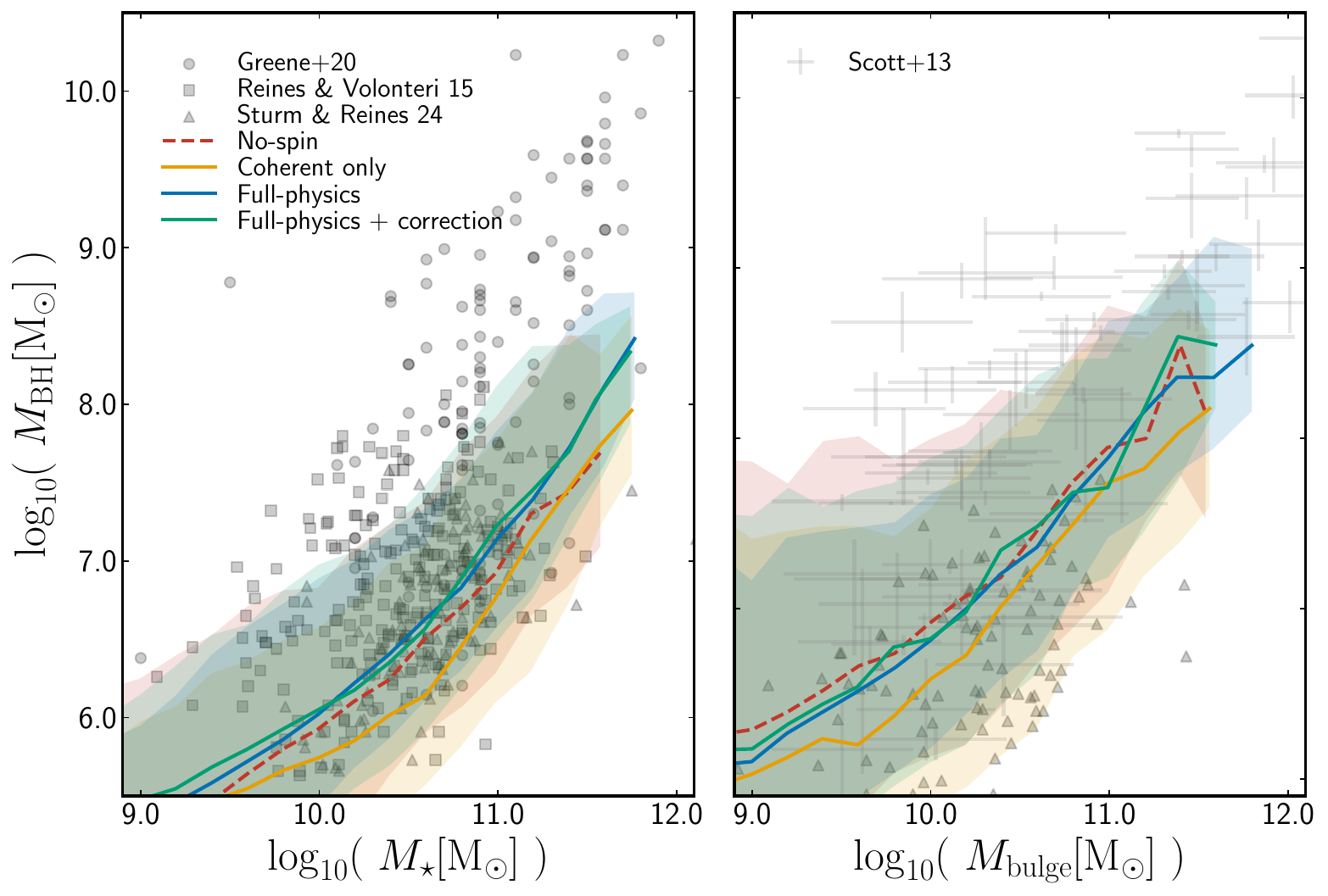}
    \caption{The $M_{\rm BH}-M_{\star}$ (left) and $M_{\rm BH}-M_{\rm bulge}$ (right) relations at $z = 0$ for all models described in Table~\ref{tab:tablemodels}. The dashed crimson line shows the \textbf{No-spin} model, solid lines are the median relation for each spin model, and the shaded regions enclose the 68th percentile range. Observational data are shown as grey symbols: circles from \citet{Greene2020}, squares from \citet{Reines&Volonteri2015}, and triangles from \citet{SturmReines2024}. The right panel additionally includes data from \citet{Scott2013THEGALAXIES}. The \textbf{Full-physics} models predict higher BH masses at fixed stellar mass than the \textbf{No-spin} model, while the \textbf{Coherent only} model shows the lowest BH masses. In the \textbf{Coherent only} model, coherent accretion drives spins to high values, raising the Novikov-Thorne radiative efficiency and reducing the fraction of accreted mass retained by the BH. Differences in the free parameter $\kappa_{\rm sink}$ between models may also contribute to the offset in BH masses at fixed stellar mass. All models remain broadly consistent with the observed scatter.}
    \label{fig:MbhMstarMbulge_z0}
\end{figure}

Figure~\ref{fig:MbhMstarMbulge_z0} shows the $M_{\rm BH}-M_{\star}$ and $M_{\rm BH}-M_{\rm bulge}$ relations at $z = 0$ for all our models alongside observational compilations. The median of all models tends to lie on the lower side of the observed relations. This is partly expected, as observational samples are flux-limited and preferentially include the most luminous and most massive BHs at fixed stellar mass. This bias shifts the observed scaling relations toward higher BH masses relative to the intrinsic population \citep{Lauer2007, Shankar2016, Barausse2017}. 

At fixed stellar mass, the \textbf{Full-physics} variants sit above the \textbf{No-spin} baseline in BH mass, while the \textbf{Coherent only} model falls below all other models. At fixed bulge mass, however, all models overlap and the differences between models largely disappear. This suggests that the $M_{\rm BH}-M_{\rm bulge}$ relation is more robust to spin model variations, as BH growth and bulge assembly are shaped by the same merger and instability events and tend to respond to AGN feedback in proportion. The offset visible in the $M_{\rm BH}-M_\star$ relation therefore arises primarily from differences in disk star formation, which is more sensitive to \textit{quasar-mode} and \textit{radio-mode} feedback variations between models, which in turn affect BH growth.

When all accretion is coherent, BHs reach systematically high spins, elevating the Novikov-Thorne radiative efficiency so that a larger share of the accreted rest-mass energy escapes as radiation rather than growing the BH. Differences between models in the calibration parameter $\kappa_{\rm sink}$ provide an additional source of spread in BH mass independent of the spin physics. For the \textbf{Full-physics} models, chaotic accretion episodes keep the average spin lower, reducing the radiative efficiency and allowing more of the infalling mass to be retained by the BH. The \textbf{Full-physics + correction} model produces more jet power than the \textbf{Full-physics} model within this stellar mass range, yet its BH masses at fixed stellar mass are comparable or slightly higher. The enhanced \textit{radio-mode} heating from the mechanical feedback floor suppresses star formation more effectively, reducing stellar masses across the population and shifting the $M_{\rm BH}-M_\star$ relation upward. 
 
Despite this offset, all models are broadly consistent with one another and with the scatter in the observed data. We note that the {\sc TNG100-1} merger trees do not adequately sample cluster-size halos, limiting our ability to populate the high-mass end of the $M_{\rm BH}-M_{\star}$ relation. 

Figure~\ref{fig:JetPower_redshift} shows the AGN energy density as a function of redshift for all our models. Across all models and redshifts, \textit{radio-mode} feedback dominates the total AGN energy budget. This is the electromagnetic jet power delivered to the hot halo via the Blandford-Znajek mechanism, which couples directly to the hot gas reservoir and drives the fountain feedback responsible for quenching massive galaxies. The \textit{quasar-mode} contribution falls around two orders of magnitude below the \textit{radio-mode} at all epochs. This reflects the fact that \textit{quasar-mode} feedback acts directly on the cold gas disk, ejecting gas to the hot halo or circumgalactic medium. Its energy output is therefore limited by the available cold gas reservoir. Over cosmic time, this reservoir is depleted by star formation and further suppressed by \textit{radio-mode} feedback, which prolongs the cooling time of the hot halo and limits the replenishment of cold gas.

The \textbf{No-spin} model consistently produces the highest total energy output at $z \lesssim 4$, driven by its fixed radiative efficiency of $\eta = 0.3$ applied to both \textit{radio-mode} and \textit{quasar-mode} feedback. Our \textbf{Full-physics + correction} model closely tracks the \textbf{Full-physics} model across all redshifts, but produces slightly higher jet power. This offset is driven by the minimum mechanical efficiency floor $\eta_{\rm floor} = 0.02$, which maintains a baseline coupling between the jet and the halo at all times. At low spins, the BZ efficiency $\eta_{\rm EM}$ would otherwise fall well below this floor, suppressing jet feedback. Among the spin models, the \textbf{Full-physics} model produces more total AGN energy than the \textbf{Coherent only} model at low redshift, despite having lower spin. This is because lower radiative losses at low spin allow more mass to be retained by the BH, driving higher BH masses and consequently higher Bondi accretion rates. However, as shown in Figure~\ref{fig:SMF_z0}, this elevated energy output does not translate into effective quenching because both $\eta_{\rm EM}$ and $\varepsilon_{\rm thin}$ are small at low spin, decoupling the feedback from the halo and cold disk respectively. 

At $z \gtrsim 4$, all models that track BH spin show an upturn in energy density from \textit{radio-mode}, likely driven by the increasing abundance of super-Eddington accreting sources at early epochs \citep{PorrasValverde2026MassAssembly}, where the MAD prescription predicts strong spin-up followed by efficient jet launching at elevated accretion rates \citep{Ricarte2023Spin}. In contrast, the \textbf{No-spin} model shows a decrease in energy output at these redshifts. As we shall see, the increased energy density of {\sc Serotina} can be explained by its spin distribution, peaked relatively higher than that of the \textbf{Full-physics} model. It is reassuring that introducing an incoherent mode in {\sc Mad Sage} moderates the energy output without completely driving all spins to $\approx$ 0.

Finally, all our models are broadly consistent with the observed kinetic energy density from \citet{Smolcic2017} at $z \lesssim 1$, which is estimated by modelling the evolution of the radio luminosity function and converting 1.4\,GHz radio luminosities to jet powers. Our models fall within or slightly above the observed band, suggesting that the predicted jet energy densities are physically plausible, though a more direct comparison requires accounting for the selection effects inherent in radio-selected AGN samples.

\begin{figure*}
    \centering
    \includegraphics[width=\linewidth]{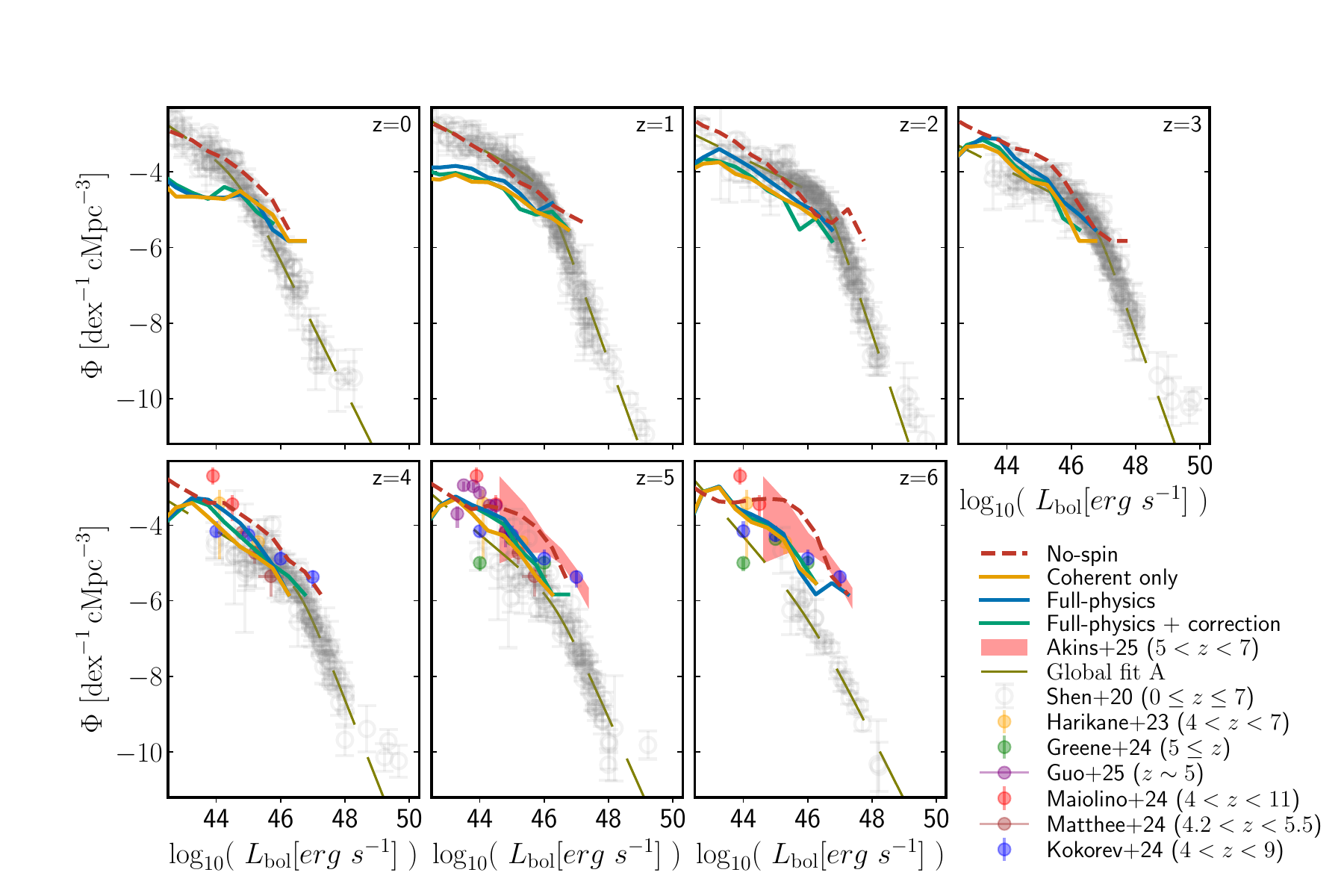}
    \caption{The AGN bolometric luminosity function from $z=0$ to $z=6$ for all our models. Observational data from \citet{Shen2020AGNbol} are shown in grey, compiled from multiple wavebands including rest-frame ultraviolet, B-band, soft and hard X-ray, and mid-infrared, with bolometric corrections applied. The dashed olive line shows Global fit A from \citet{Shen2020AGNbol}. At $z \geq 4$, additional data are included from \citet{Harikane+2023} ($4 < z < 7$), \citet{Maiolino+2024} ($4 < z < 11$), \citet{Matthee+2024} ($4.2 < z < 5.5$), and \citet{Kokorev+2024} ($4 < z < 9$). At $z = 5-6$, the shaded red region shows the inferred luminosity function assuming that the Little Red Dots identified by JWST are AGNs \citep{Akins+2025}, with further constraints from \citet{Greene+2024} ($z \geq 5$) and \citet{Guo+2025} ($z \sim 5$). The BH spin models remain mutually consistent across the full redshift range. The \textbf{No-spin} model overpredicts the AGN bolometric luminosity function at $z \gtrsim 4$, with its bright-end abundance more closely matching the JWST-inferred excess of luminous AGN. The BH spin models also overlap with the JWST constraints at those redshifts, but predict lower number densities that are more consistent with \citet{Shen2020AGNbol}.}
    \label{fig:AGNbol_z0z6}
\end{figure*}

Figure~\ref{fig:AGNbol_z0z6} shows the AGN bolometric luminosity function from $z=0$ to $z=6$. At $z \sim 2$-$3$, all models reproduce the \citet{Shen2020AGNbol} constraints well across the full luminosity range. Above $z \sim 4$, the \textbf{No-spin} model overpredicts the number density of luminous AGN relative to the spin models, with its bright end tracing the JWST-inferred excess from \citet{Akins+2025} more closely. The three BH spin models remain tightly clustered with one another at all redshifts. At $z \geq 4$ they overlap with the JWST observations and stay in close agreement with \citet{Shen2020AGNbol}, predicting lower number densities than the \textbf{No-spin} model. At $z \leq 1$, all spin models underpredict the faint end of the luminosity function ($\log_{10} L_{\rm bol} \lesssim 44$), whereas the \textbf{No-spin} model reproduces it more successfully. This faint-end deficit in the spin models is consistent with the broader picture emerging from this work. At sub-Eddington accretion rates, BHs undergo net spin-down through the electromagnetic torque, reducing the radiative efficiency over time and suppressing the bolometric output at low redshift, where most BHs accrete well below the Eddington limit.

\begin{figure}
    \centering
    \includegraphics[width=\linewidth]{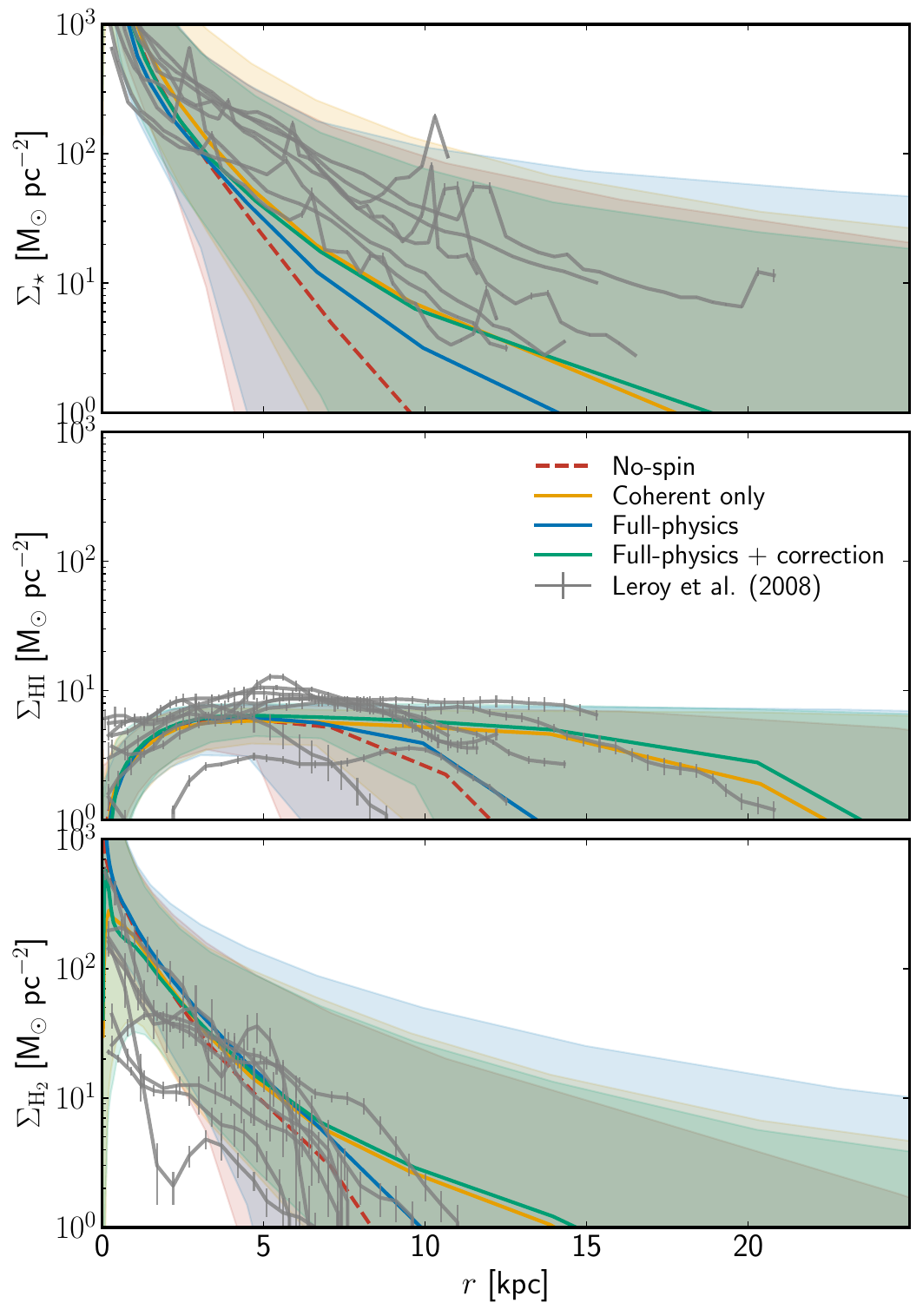}
    \caption{Stellar (top), H\,{\sc i} (middle), and H$_2$ (bottom) surface density profiles at $z = 0$ for all our models. Dashed and solid lines show the median profile and shaded regions enclose the 68th percentile range across central galaxies with stellar mass $M_\star > 10^{10}\,\mathrm{M}_\odot$, cold gas mass $M_{\rm cold} > 10^{9.2}\,\mathrm{M}_\odot$, and bulge-to-total ratio $B/T < 0.5$. Observed profiles of galaxies with similar properties from \citet{Leroy2008} are shown in grey. All models broadly reproduce the observed surface density profiles at small radii ($r \lesssim 5\,\mathrm{kpc}$). At larger radii, the \textbf{No-spin} and \textbf{Full-physics} models predict more compact profiles than the \textbf{Coherent only} and \textbf{Full-physics + correction} models.}
    \label{fig:Surfacedensityprof_z0}
\end{figure}

Figure~\ref{fig:Surfacedensityprof_z0} shows the stellar, H\,{\sc i}, and H$_2$ surface density profiles at $z = 0$ for all our models. All models converge at small radii ($r \lesssim 5\,\mathrm{kpc}$), indicating that the inner disk structure is relatively insensitive to the choice of AGN feedback prescription. This is noteworthy given that \textit{quasar-mode} feedback is the AGN channel directly coupled to the cold gas disk, ejecting gas from the inner to the outer regions depending on the energy carried by the outflow. At fixed galaxy mass, the \textit{quasar-mode} energy in the \textbf{Coherent only} model closely follows the \textbf{Full-physics} model up to $M_\star \sim 10^{11.5}\,\mathrm{M}_\odot$, above which it rises sharply matching the \textbf{Full-physics + correction} model. As a function of BH mass, the \textbf{Coherent only} model traces the \textbf{Full-physics + correction} model (see Appendix~\ref{app:AGNenergy_logTSM}).

At larger radii, the \textbf{Full-physics} model predicts galaxy sizes remarkably similar to the \textbf{No-spin} model, despite the inclusion of self-consistent spin evolution. In general, the two models produce similar disk sizes at large radii, despite their different AGN prescriptions. Conversely, the \textbf{Coherent only} and \textbf{Full-physics + correction} models both impose stronger effective AGN feedback, either through purely coherent spin-up or through the energy floor.

We note that this pattern in the outer-disk size does not track the total AGN energy output of each model in a simple way (Figure~\ref{fig:JetPower_redshift}). The \textbf{No-spin} model, which injects the most AGN energy of all four models, produces the most compact disks. Among the spin-dependent models, however, neither the model with the least jet power (\textbf{Coherent only}) nor the model with the most (\textbf{Full-physics + correction}) shows a correspondingly compact profile. Both produce the most extended disks. 

We caution against over-interpreting this as evidence that AGN feedback plays no direct role in setting the outer-disk profile. \textit{Quasar-mode} feedback is physically coupled to the cold gas disk and can expel gas from the inner disk outward depending on the energy carried by the outflow. Some genuine dependence of the radial profile on AGN energy is expected on physical grounds. Rather, we suggest that the freely calibrated $E_{\rm SN}$ parameter, which differs between models (Table~\ref{tab:tablemodels}), may be partially erasing this underlying relationship. Because $E_{\rm SN}$ and $\kappa_{\rm sink}$ (the amount of gas transferred between annuli in the disk) are used jointly for calibration, any radial signature imprinted by AGN feedback on the disk may be partially obscured by the supernova calibration. This overlap is made harder to disentangle by the higher supernova feedback energy in the \textbf{Coherent only} and \textbf{Full-physics + correction} models relative to the \textbf{No-spin} and \textbf{Full-physics} models. Yet the profiles of all BH spin models remain consistent with the observed profiles of \citet{Leroy2008}.

\subsection{Population-level spin predictions}\label{sec:spin_results}

\begin{figure*}
    \centering
    \includegraphics[width=\linewidth]{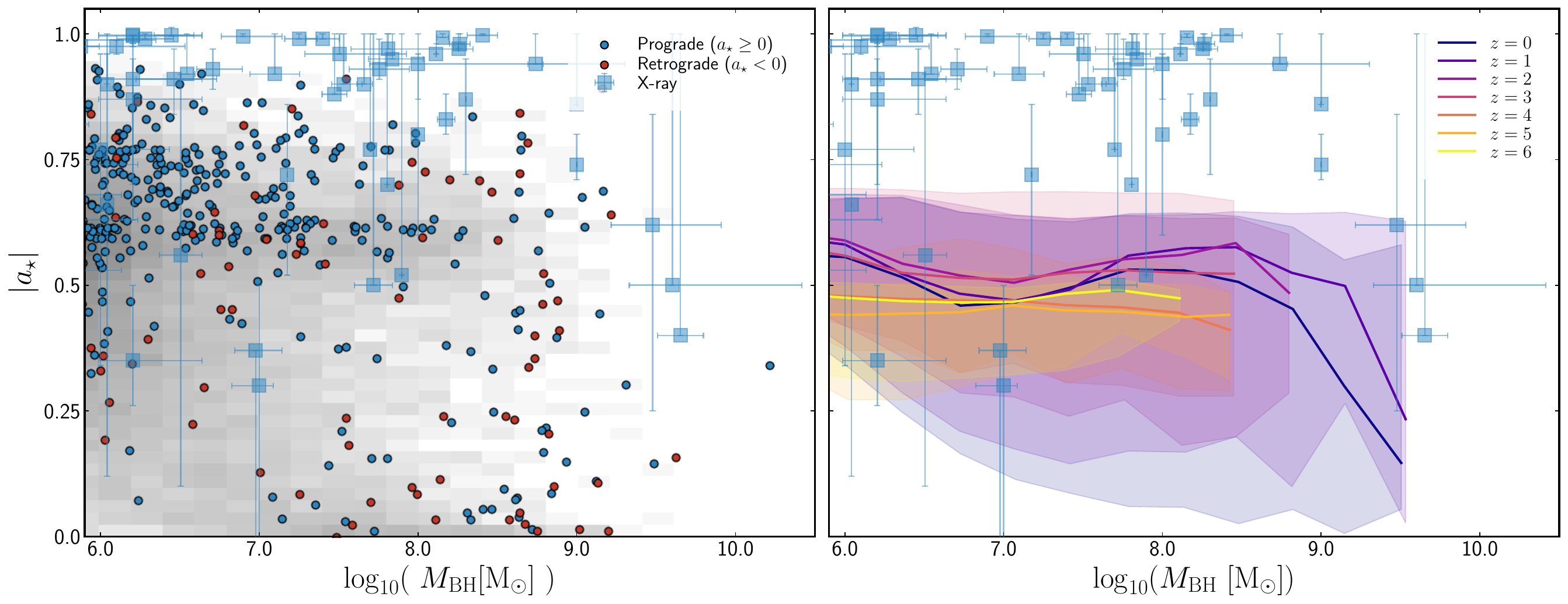}
    \caption{\textit{Left:} Absolute magnitude of the BH spin as a function of BH mass for our \textbf{Full-physics + correction} model. The 2D histogram in gray shows the full population at $z=0$. Blue and red circles highlight actively accreting BHs ($f_{\rm Edd} > 0.01$) with prograde ($a_\star \geq 0$) and retrograde ($a_\star < 0$) spin respectively. The majority of prograde BHs cluster at $|a_\star| \sim 0.5$-$0.7$, while retrograde BHs have wide spread of high and low spin magnitudes. Blue squares with error bars show BH spin measurements from X-ray reflection spectroscopy compiled in \citet{Ricarte+2025}. \textit{Right:} Median $|a_\star|-M_{\rm BH}$ relation from $z=0$ (dark blue) to $z=6$ (yellow), with shaded regions enclosing the 68th percentile range at each epoch. X-ray observations are shown for comparison. At low redshift, the median spin decreases with increasing BH mass. Massive BHs are systematically spun down by repeated episodes of Bondi accretion. At higher redshift ($z \gtrsim 3$), the median spin is broadly flat across BH mass, pointing to a regime in which cold disk feeding dominates BH growth at all masses.}
    \label{fig:BHmassBHspin_z0z6}
\end{figure*}

We now examine how the self-consistent coupling between galaxy-scale gas accretion and the MAD jet mechanism shapes the predicted SMBH spin distribution across BH mass and cosmic time. Figure~\ref{fig:BHmassBHspin_z0z6} shows the absolute magnitude of the BH spin as a function of BH mass for our \textbf{Full-physics + correction} model. At low BH mass ($M_{\rm BH} \lesssim 10^7\,\mathrm{M}_\odot$), the majority of BHs have spin magnitudes in the range $|a_\star| \sim 0.5$-$0.7$, modestly lower than the predictions of the Fiducial \textsc{Serotina} semi-analytic model \citep{Ricarte+2025} due to the addition of an incoherent accretion mode. Among actively accreting BHs ($f_{\rm Edd} > 0.01$), prograde spins cluster in the range $|a_\star| \sim 0.5$-$0.7$ , while retrograde spins have a wide range of values ($|a_\star| \sim 0-0.8$). The concentration of spins near $|a_\star| \sim 0.5$-$0.7$ is the result of coherent accretion driving spin-up and the MAD jet mechanism driving spin-down, with the near-Eddington equilibrium spin settling near $a_\star \approx 0.7$ \citep{Ricarte2023Spin}. Lower-mass BHs retain higher spins, consistent with being dominated by coherent disk accretion at higher Eddington ratios. This is in line with \citet{PorrasValverde2026MassAssembly}, who showed that BHs with $M_\bullet \sim 10^6\,\mathrm{M}_\odot$ grow predominantly through secular disk instabilities in {\sc Dark Sage} at all redshifts. Super-Eddington accretion rates peak at high redshift and originate from the instabilities channel.

At low redshift ($z<3$), the median spin decreases with increasing BH mass. The most massive BHs ($M_{\rm BH} \gtrsim 10^8\,\mathrm{M}_\odot$) reach the lowest spin magnitudes, reflecting spin-down driven by prolonged incoherent Bondi accretion from the hot halo. This is consistent with earlier semi-analytic predictions showing that randomly oriented accretion episodes drive spins toward $|a_\star| \sim 0.1$-$0.3$, and that BH-BH mergers are unlikely to produce large remnant spins unless the progenitor spins are strongly aligned with the orbital angular momentum \citep{Volonteri2005, Berti2008}. At high redshift ($z>3$), the median spin remains broadly constant across BH mass. Lower-mass BHs retain higher spins, consistent with being dominated by coherent disk accretion at super-Eddington rates.

We compare against BH spin measurements from X-ray reflection spectroscopy compiled in \citet{Ricarte+2025} obtained from \citet{Bambi2021, Reynolds2021, Walton2021, Sisk-Reynes2022, Mallick2022}. Note that these measurements carry significant model uncertainties and selection effects \citep{Ricarte+2025}. In particular, X-ray reflection spectroscopy assumes the existence of a geometrically thin, radiatively efficient accretion disk extending to the innermost stable circular orbit, which applies to $f_{\rm Edd} \in [0.01, 1]$ \citep{Reynolds2021}. To approximately match these selection effects, we restrict the red and blue points in Figure~\ref{fig:BHmassBHspin_z0z6} to actively accreting BHs with $f_{\rm Edd} > 0.01$. The observed distribution peaks near $|a_\star| \approx 1$ but extends to significantly lower spin values, which our model broadly reproduces for the actively accreting population.

\begin{figure}
    \centering
    \includegraphics[width=\linewidth]{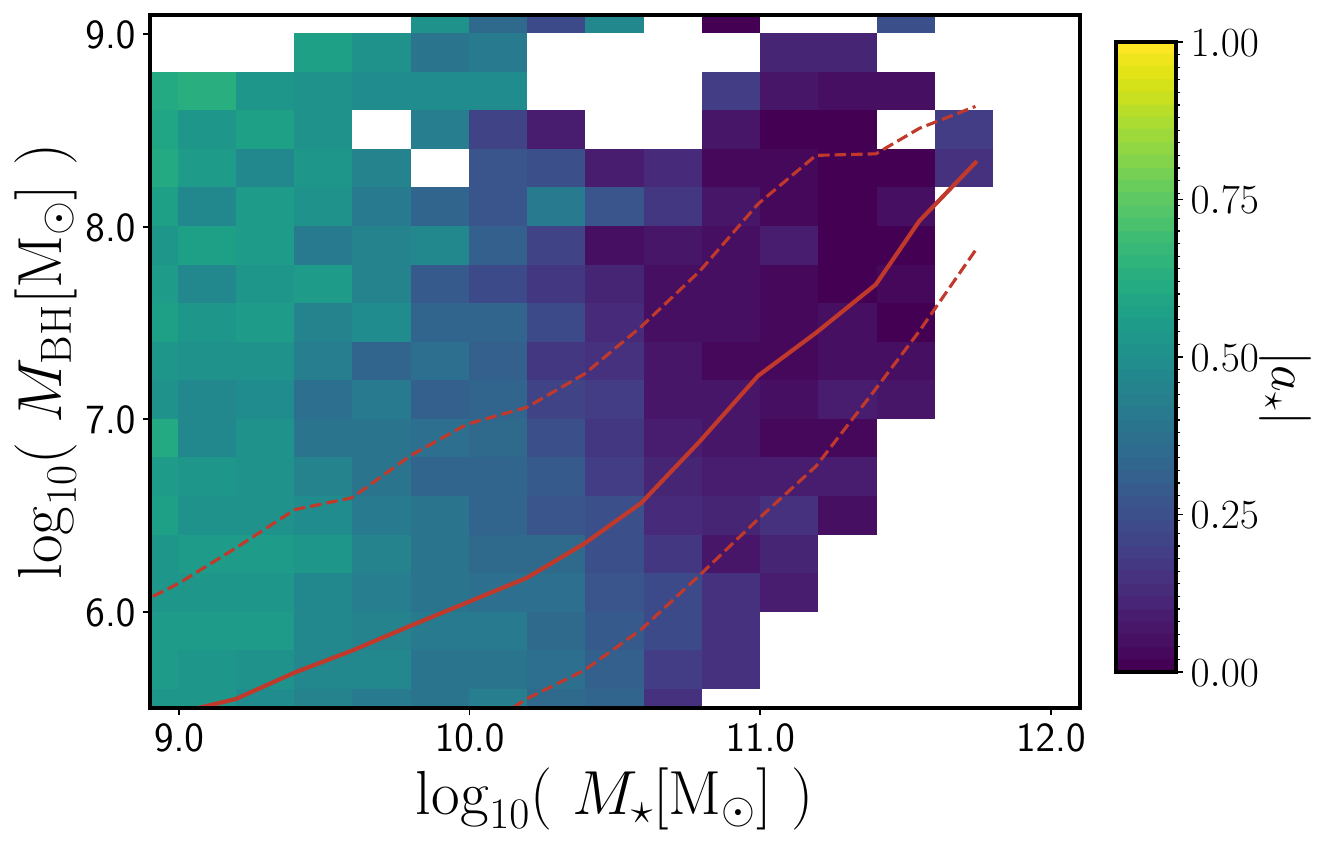}
    \caption{BH mass-stellar mass relation colored by the median absolute BH spin at $z = 0$ for our \textbf{Full-physics + correction} model. Each bin shows the median $|a_\star|$ for galaxies within that cell. The red solid line shows the median $M_{\rm BH}-M_\star$ relation, with dashed lines enclosing the 68th percentile ranges. At fixed BH mass, spin tends to decrease with increasing stellar mass, reflecting the shift from disk-fed accretion in low-mass galaxies towards incoherent Bondi accretion from the hot halo in more massive systems, where the Blandford-Znajek mechanism efficiently extracts spin energy to power the jet, and the BH-BH mergers with isotropically distributed progenitor spins further reduce or enhance the remnant spin magnitude.}
    \label{fig:MbhMstarcoloredBHspin_z0}
\end{figure}

\begin{figure*}
    \centering
    \includegraphics[width=\linewidth]{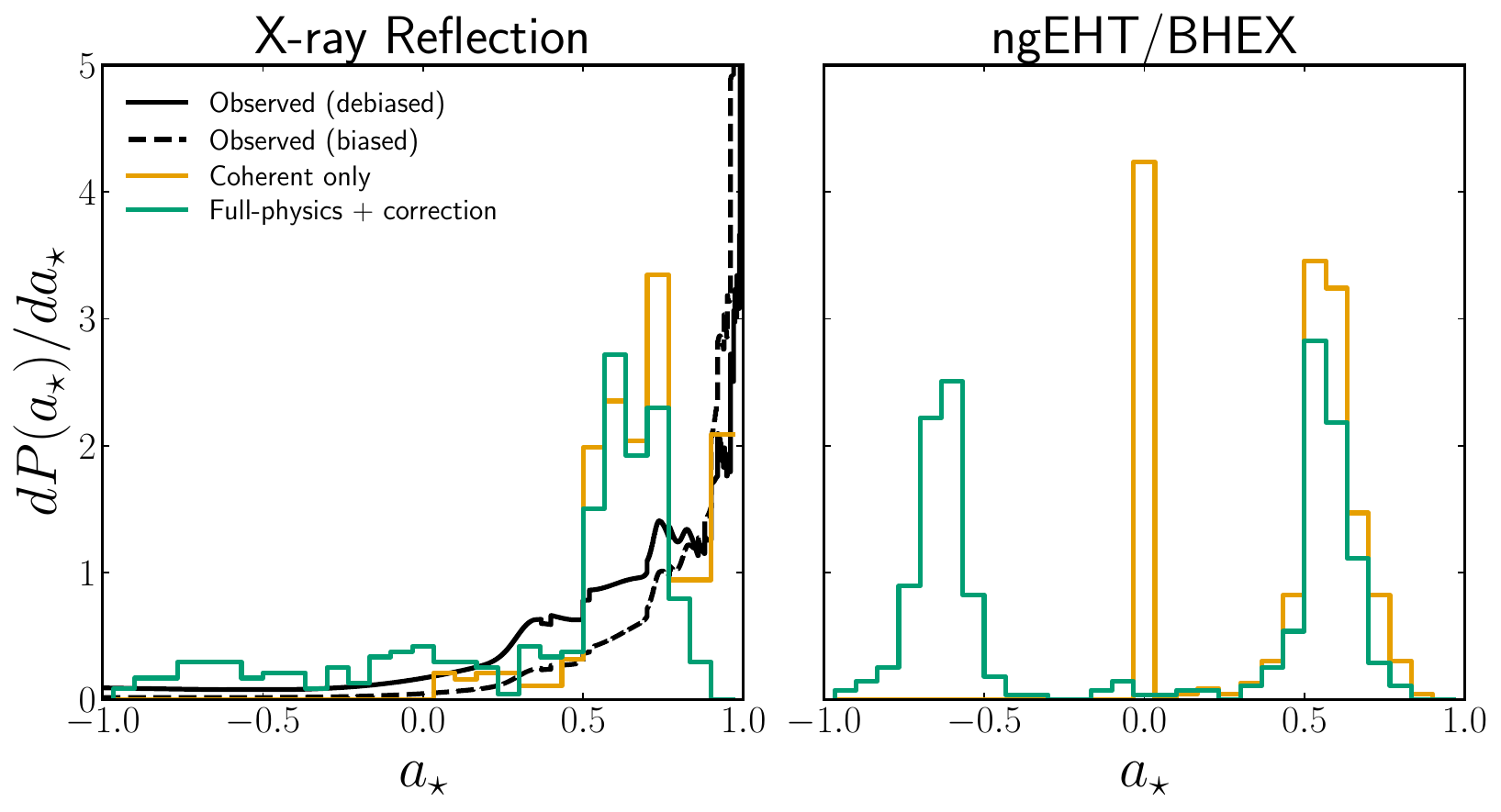}
    \caption{Normalized BH spin distributions for our spin models compared against observations at $z=0$. \textit{Left:} Distributions for BHs observable with X-ray reflection spectroscopy, selecting actively accreting BHs with $f_{\rm Edd} > 0.01$. The black solid and dashed lines show the observed debiased and biased spin distributions compiled in \citet{Ricarte+2025}. The two {\sc Mad Sage} models shown, the \textbf{Coherent only} model and the \textbf{Full-physics + correction} model, both predict spin distributions peaked at moderate values ($a_\star \sim 0.5$-$0.7$), falling short of both the biased and debiased observed distributions which peaks near $|a_\star| \approx 1$. \textit{Right:} Distributions of BHs detected by the next-generation Event Horizon Telescope and/or the Black Hole Explorer, selecting massive, quiescent BHs with $M_{\rm BH} > 10^8\,\mathrm{M}_\odot$ and $f_{\rm Edd} < 10^{-4}$. In both panels, the orange and green lines show our \textbf{Coherent only} and \textbf{Full-physics + correction} models respectively. {\sc Mad Sage} models predict bimodal spin distributions for massive quiescent BHs, with populations at both moderate and low spin magnitudes driven by the balance between coherent disk feeding and prolonged ADAF-dominated spin-down.}
    \label{fig:BHspin_Xref_ngEHT}
\end{figure*}

To further understand how the spin distribution depends on the host galaxy properties, Figure~\ref{fig:MbhMstarcoloredBHspin_z0} maps the median spin across the BH mass-stellar mass plane at $z = 0$. Figure~\ref{fig:MbhMstarcoloredBHspin_z0} displays the BH mass-stellar mass relation colored by the median absolute BH spin $|a_\star|$ at $z = 0$ for our \textbf{Full-physics + correction} model. At low stellar masses, BHs tend to have spins of $|a_\star| \gtrsim 0.5$. This trend reflects the shift in the dominant accretion channel across the galaxy mass function. Low-mass galaxies are predominantly disk-dominated, with coherent gravitational instability accretion efficiently spinning up their central BHs. Massive galaxies are more gas-poor and their BHs accrete primarily through the incoherent Bondi channel from the hot halo. This drives spin-down through repeated accretion events with randomly oriented angular momenta \citep{KingPringle2006, King2008, Barausse2012}. In the most massive systems, prolonged ADAF-dominated accretion and spin energy extraction via the Blandford-Znajek mechanism together drive spins toward the low equilibrium value $a_\star \approx 0.035$ set by the MAD jet mechanism \citep{Narayan2022, Ricarte2023Spin}.

\subsection{Predicted spin distributions across observational windows}\label{sec:BHspin_observable}

We now translate our model spin distributions into predictions tailored to the selection effects of current and future observational facilities, focusing on X-ray reflection spectroscopy, the ngEHT and BHEX, and LISA, each of which probes a distinct population of BHs across different mass, Eddington ratio, and redshift regimes.

\begin{figure*}
    \centering
    \includegraphics[width=\linewidth]{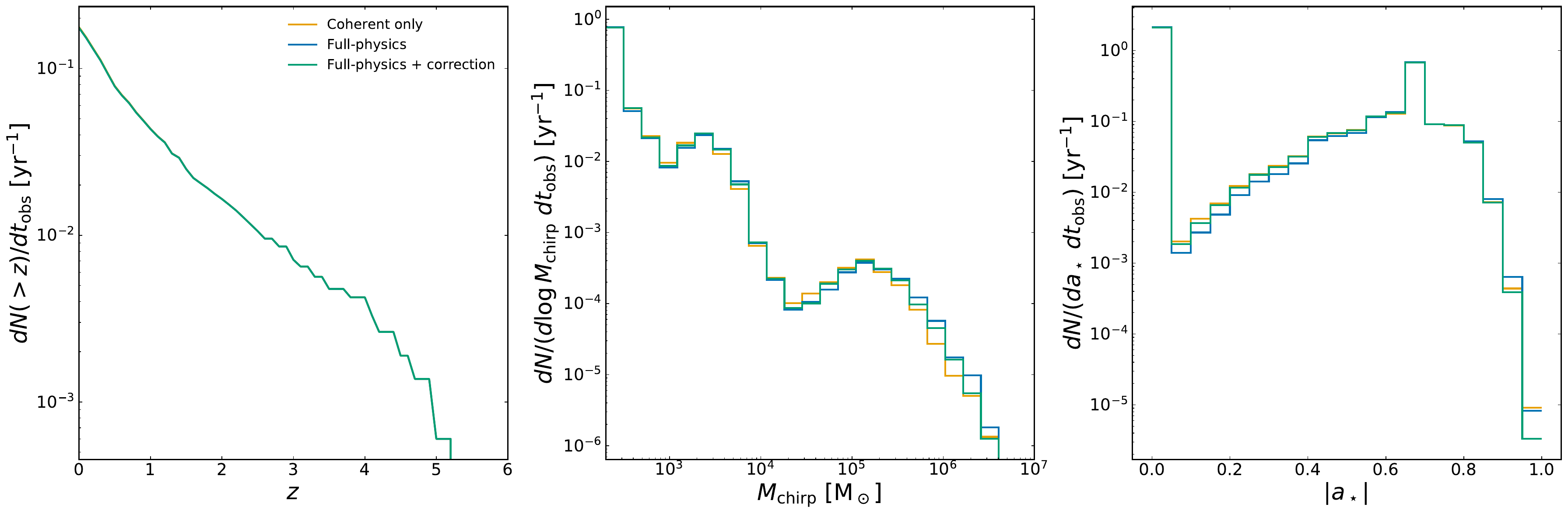}
    \caption{Predicted LISA BH-BH merger event rates for three of our spin models, showing the cumulative event rate as a function of redshift (\textit{left}), the differential event rate per logarithmic chirp mass interval (\textit{center}), and the differential event rate per unit remnant spin magnitude (\textit{right}). All rates are expressed per observer-frame year and include the $(1+z)^{-1}$ time-dilation correction. Events are selected within the LISA sensitivity band ($10^4 \lesssim M_{\rm total}/\mathrm{M}_\odot \lesssim 10^{10}$, $z \leq 10$) using a SNR interpolation table. {\sc Mad Sage} models all use the same {\sc TNG100-1} merger trees and a BH is seeded for every halo. Therefore, all three models produce nearly identical event rates as a function of redshift and chirp mass, since they share the same \textsc{TNG100-1} merger trees and BH seeding prescription. The remnant spin distributions are indistinguishable between models, all peaking near $|a_{\star,\rm rem}| \approx 0.7$.}
    \label{fig:LISA_BHspins}
\end{figure*}

Figure~\ref{fig:BHspin_Xref_ngEHT} shows the normalised BH spin distributions for two of our spin models, the \textbf{Coherent only} model and the \textbf{Full-physics + correction} model, compared against observational data, applying selection cuts designed to match the detection biases. The left panel corresponds to BHs detectable via X-ray reflection spectroscopy, selecting actively accreting BHs with $f_{\rm Edd} \in [0.01, 1]$ to match the underlying assumption of a radiatively efficient thin disk \citep{Reynolds2021}. We compare against the observed biased and debiased spin distributions compiled in \citet{Ricarte+2025} from 55 measurements in the literature. Both distributions peak near $|a_\star| \approx 1$ with a broad tail towards more moderate spin values. 

Both {\sc Mad Sage} models predict spin distributions peaked at moderate values ($a_\star \sim 0.5$-$0.7$) though falling short of both the biased and debiased observed distributions at the highest spin. The \textbf{Coherent only} model produces a narrower distribution skewed toward higher spins than the \textbf{Full-physics + correction} model, a consequence of the systematic spin-up driven by purely coherent accretion. We note that the observed distributions carry significant model uncertainties and selection effects that discourage more quantitative comparison.

The right panel corresponds to BHs observable by the ngEHT and/or BHEX \citep{Pesce2021, Johnson2024, Zhang2025ApJngEHT}, selecting massive quiescent BHs with $M_{\rm BH} > 10^8\,\mathrm{M}_\odot$ and $f_{\rm Edd} < 10^{-4}$. Spin measurements from ngEHT/BHEX will probe a fundamentally different population than X-ray reflection spectroscopy, selecting for nearby resolved sources at low Eddington ratios where the accretion flow is optically thin enough to reveal the BH shadow \citep{Yuan&Narayan2014, EventHorizonTelescopeCollaboration2019ApJ, EventHorizonTelescopeCollaboration2022ApJ}. In this regime, ADAF-dominated accretion is expected to govern the spin evolution \citep{Narayan2022, Ricarte+2025}.

Both models predict bimodal spin distributions for this population, with populations at both moderate and low spin magnitudes. The \textbf{Full-physics + correction} model predicts retrograde spins peaking at $a_\star \approx -0.7$ and prograde spins peaking at $a_\star \approx 0.5$, reflecting BHs whose spin orientation has been randomised by prolonged incoherent Bondi accretion while retaining moderate spin magnitudes from earlier coherent disk feeding. The \textbf{Coherent only} model produces a population at $a_\star \approx 0.5$ alongside a second population driven toward low spin magnitudes ($a_\star \approx 0$), consistent with ADAF-dominated accretion bringing the most quiescent systems towards the MAD equilibrium value of $a_\star \approx 0.035$.

Our results are consistent with {\sc Serotina}, which finds a population at moderate spin magnitudes with some retrograde BHs arising from its disruptive merger prescription \citep{Ricarte+2025}. This picture is further supported by cosmological simulations that find an increasing fraction of retrograde systems toward higher BH masses \citep{BustamanteVolker2019, Sala2024}.

Beyond electromagnetic observations, gravitational wave detectors offer an independent and complementary window onto BH spin through the remnant spins of merging BH binaries, which we now explore in the context of LISA. Figure~\ref{fig:LISA_BHspins} shows the predicted LISA BH-BH merger event rates for our spin models. To evaluate LISA detectability, we construct a signal-to-noise ratio (SNR) interpolation table over a three-dimensional grid of total BH mass, mass ratio, and redshift using the analytic LISA sensitivity curve of \citet{Robson2019} and the stationary phase approximation for circular inspiral waveforms. The SNR is evaluated at each grid point by integrating the squared strain amplitude over the LISA frequency band, accounting for the cosmological time dilation of the source-frame merger time. Events with $\mathrm{SNR} > 8$ are classified as detectable.

\begin{figure}
    \centering
    \includegraphics[width=\linewidth]{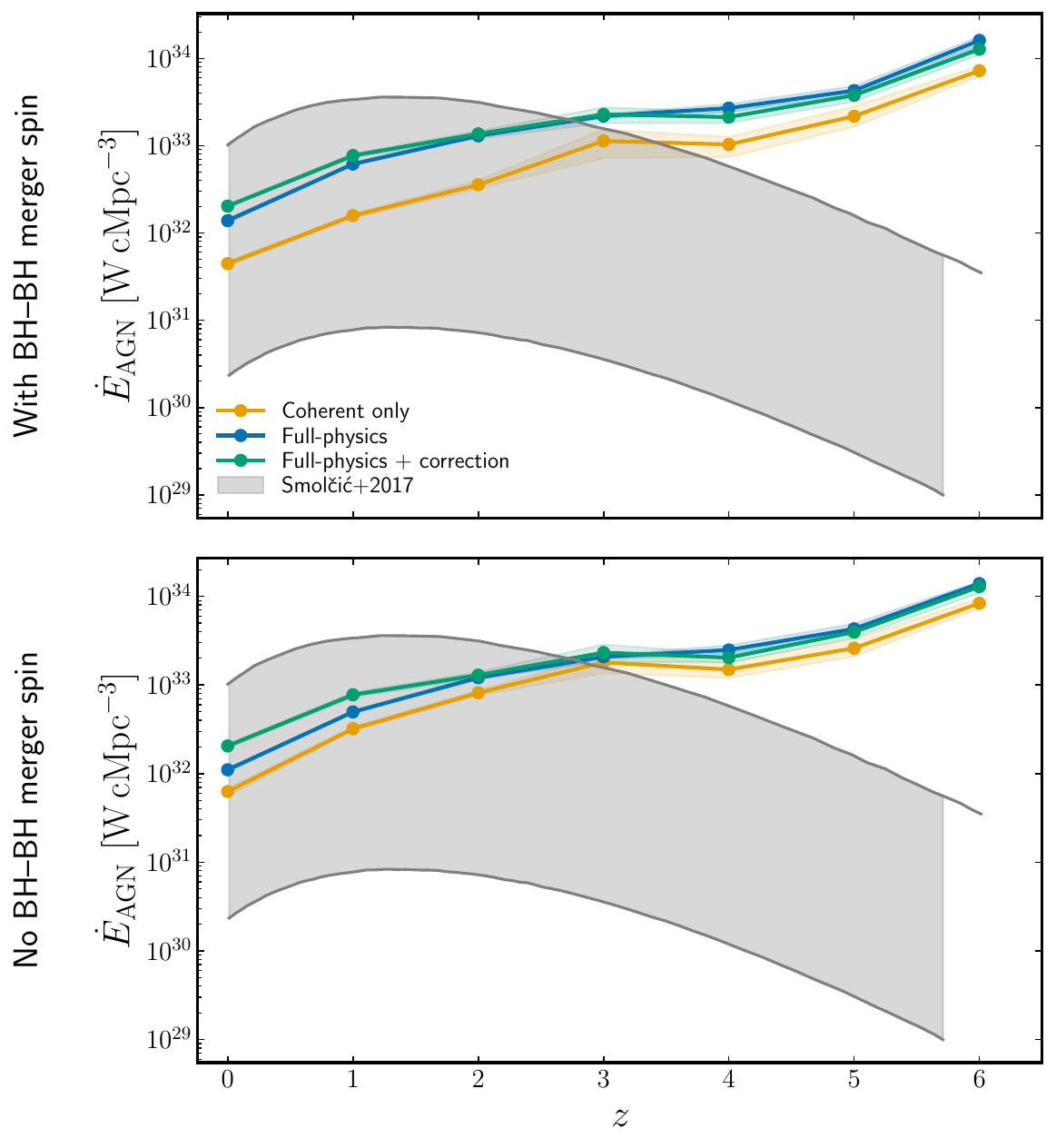}
    \caption{The total AGN energy density as a function of redshift for models with BH-BH merger spin (\textit{top}) and without BH-BH merger spin (\textit{bottom}). The grey shaded region displays the observed kinetic energy density from radio-selected AGN compiled by \citet{Smolcic2017}. Shaded bands around the model lines show the 68th percentile bootstrap uncertainty from the finite simulation volume. For the \textbf{Full-physics} model, BH-BH merger spin slightly increases the total AGN energy density, whereas for \textbf{Coherent only} it slightly decreases it. The \textbf{Full-physics + correction} model remains unaffected.}\label{fig:AGNenergy_BHspinmerger_nomerger}
\end{figure}

All three {\sc Mad Sage} models shown in Figure~\ref{fig:LISA_BHspins} produce nearly identical event rates as a function of redshift and chirp mass, a direct consequence of sharing the same {\sc TNG100-1} merger trees and the same BH seeding prescription, which is dependent on halo mass. The chirp mass distribution has a tri-modal structure. The first peak near $M_{\rm chirp} \sim 10^2\,\mathrm{M}_\odot$ corresponds to the seed-mass BHs merging before significant growth occurs. The intermediate peak near $10^3$-$10^4\,\mathrm{M}_\odot$ arises from BHs with stagnant growth, either those that underwent a brief super-Eddington episode at high redshift and later became inactive, or those that have grown steadily but slowly since early epochs. The high-mass peak near $10^5$-$10^6\,\mathrm{M}_\odot$ corresponds to BHs that experienced sustained super-Eddington accretion at high redshift and continued growing. The bulk of detectable mergers occurs at $z \lesssim 3$, consistent with the peak of BH merger activity across cosmic time. 

The remnant spin distribution remain consistently similar between models. All BH spin models predict a peaks near $|a_{\star,\rm rem}| \approx 0.7$ and a common low-spin population near $|a_{\star,\rm rem}| \approx 0.05$, arising from either sub-Eddington accretors and/or BH-BH mergers between progenitors with randomly oriented spins. 

\begin{figure}
    \centering
    \includegraphics[width=\linewidth]{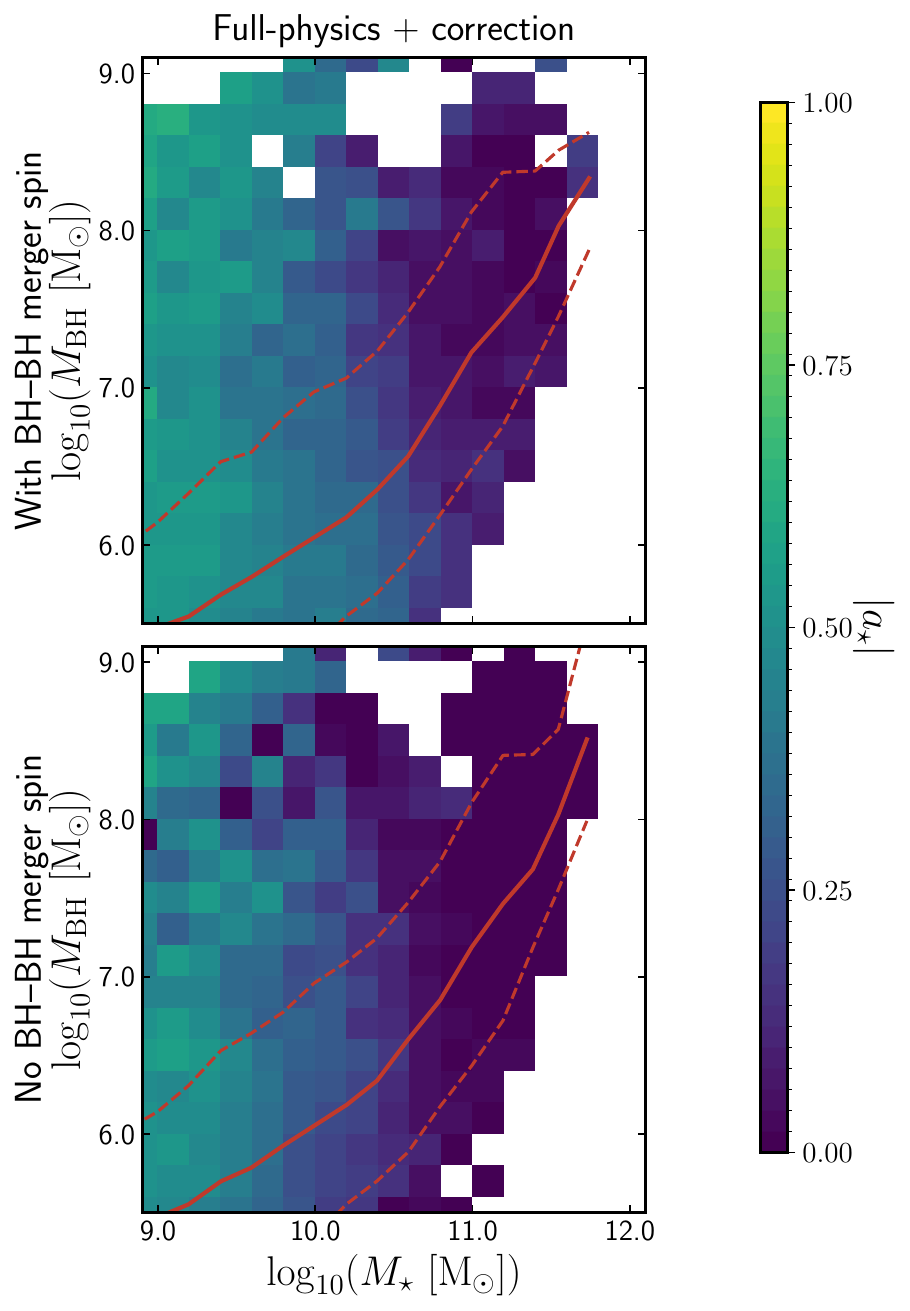}
    \caption{BH mass-stellar mass relation colored by the median absolute magnitude of BH spin at $z=0$ for the \textbf{Full-physics + correction} model with BH-BH merger spin (\textit{top}) and without BH-BH merger spin (\textit{bottom}). Each bin shows the median $|a_\star|$ for galaxies within that cell, requiring at least five objects per bin. The red solid line shows the median $M_{\rm BH}-M_\star$ relation and the dashed lines enclose the 68th percentile range. The \textbf{Full-physics + correction} prescription shows the same spin distribution irrespective of the BH-BH merger spin treatment. At fixed BH mass, $|a_\star|$  decreases with increasing stellar mass.}\label{fig:heatmap_BHspinmerger_nomerger}
\end{figure}

\subsection{The effect of BH-BH merger spin}\label{sec:BH-BHmergerspin}

Having established the baseline predictions of our spin models, we now isolate the effect of BH-BH merger spin by comparing otherwise identical models with and without the \citet{Rezzolla2008} remnant spin prescription. Figure~\ref{fig:AGNenergy_BHspinmerger_nomerger} compares models with and without BH-BH merger spin to isolate the effect of the merger spin prescription on the galaxy population and AGN energy output. The effect of BH-BH merger spin on the total AGN energy density at low redshift ($z \lesssim 2$) depends on the accretion prescription. For \textbf{Full-physics}, BH-BH mergers preferentially spin up BHs near the ADAF equilibrium ($a_\star \approx 0.035$), producing remnants with moderate spin $|a_{\star, \rm rem}| \sim 0.5$-$0.7$ that significantly increase $\eta_{\rm EM} \propto a_\star^2$ and the resulting jet power. For \textbf{Coherent only}, where BHs are already driven to high spins by coherent accretion, BH-BH mergers instead act as a spin-down mechanism, reducing the AGN energy output. \textbf{Full-physics + correction} is unaffected by the BH-BH merger spin prescription. The most massive BHs, which are spun up by mergers, do not dominate the total AGN energy output at low redshift. For low-mass BHs that could otherwise be spun down, the imposed $\eta_{\rm floor}$ prevents their radiative efficiency from dropping further, leaving the AGN energy density unchanged.

The distribution of spin across the BH mass-stellar mass plane provides a complementary view of this effect, directly revealing where in the galaxy population mergers leave the strongest imprint on the spin distribution. Figure~\ref{fig:heatmap_BHspinmerger_nomerger} shows how the median BH spin varies across the BH mass-stellar mass plane split by whether BH-BH merger spin is included. We find a pronounced decrease in BH spin toward massive galaxies, where BZ spin-down and incoherent Bondi accretion dominate and drive spins toward the ADAF equilibrium value. The most notable difference introduced by BH-BH merger spin is an increase in median spin at the high BH masses. This is consistent with the higher AGN energy output seen at low redshift in the \textbf{Full-physics} models that include BH-BH merger spin (Figure~\ref{fig:AGNenergy_BHspinmerger_nomerger}), where the spin-up from mergers enhances $\eta_{\rm EM} \propto a_\star^2$ and consequently the jet power. The \textbf{Coherent only} and \textbf{Full-physics} prescriptions show qualitatively similar trends across the BH mass-stellar mass plane, with the same high-mass spin-up feature when BH-BH merger spin is included.

\section{Discussion}
\label{S:discussion}

\subsection{Limitations of the MAD prescription}
A fundamental assumption underlying {\sc Mad Sage} is that accreting BHs are always in the magnetically arrested state, with the magnetic flux at the horizon fully saturated. Support for this assumption comes from horizon-scale polarimetric observations of M87* and Sgr A*, whose measured magnetic field structures are better described by MAD than by SANE accretion models \citep{EventHorizonTelescopeCollaboration2021, EHTMW_5_2022, Wielgus2022}. Numerical simulations that follow magnetic field transport from the large-scale stellar wind environment down to the horizon of Sgr A* also find that the MAD state is reached without requiring fine-tuned initial conditions \citep{Ressler2020, Cho+2023, Cho+2024, Cho2025, Cho2026}. However, both lines of evidence apply specifically to low Eddington ratio systems, and it remains unclear whether magnetic saturation is maintained across the broad range of accretion rates and BH masses sampled by our galaxy population. 

At super-Eddington rates, there is evidence that MAD conditions operate in at least a subset of radio-loud systems \citep{Zamaninasab2014}. Also, MADs have been proposed to explain the small subset of tidal disruption events that launch relativistic jets \citep{Tchekhovskoy2014, CurdNarayan2019}, but these constitute only $\sim 1\%$ of known tidal disruption events \citep{TeboulMetzger2023} and their accretion physics may differ from the more typical super-Eddington flows in our model. The transition between MAD and SANE regimes is likely gradual \citep{Narayan2012}, and incorporating a smooth, accretion-rate dependent magnetization would improve our predictions particularly at high Eddington ratios, where the MAD assumption is most uncertain. 

\subsection{Additional mechanical feedback and the radio-loud/radio-quiet dichotomy}

The need for an additional mechanical feedback component, represented by $\eta_{\rm floor}=0.02$, to reproduce the observed stellar mass function raises a broader question about the relative contributions of spin-powered jets and accretion-powered outflows to AGN feedback. The Blandford-Znajek mechanism \citep{Blandford1977} extracts rotational energy from the BH spin via large-scale magnetic fields threading the horizon, with jet efficiency scaling as $\eta_{\rm EM}
\propto a_\star^2$. However, alternative mechanisms exist to power outflows, including the Blandford-Payne process \citep{Blandford1982} which drives magnetocentrifugal winds from the accretion disk itself, with power coupled to the disk accretion rate rather than to BH spin. GRMHD simulations of ADAF flows around non-spinning BHs find that even in the absence of spin, the system still produces outflows with an efficiency of $\eta \approx 0.02$ \citep{Cho+2023, Cho+2024}, likely caused by magnetic reconnection-driven convection \citep{Cho+2023} and disk-powered winds carrying away angular momentum \citep{Chatterjee2022}.

Observationally, \citet{vanVelzenFalcke2013} found a tight linear correlation between jet power and disk luminosity in a sample of 763 Fanaroff-Riley II quasars. The scatter attributable to internal parameters such as spin contributes at most 0.24 dex. They conclude that the power output from the inner disk likely dominates over the BZ spin extraction. This picture is consistent with the finding of \citet{Done2026} that the radio jet in both stellar and SMBHs couples to the hot accretion flow rather than to the optically thick disk. In that framework, the majority of radio-quiet AGN have jet powers consistent with a constant fraction of the accretion power, implying low to moderate spins across this population. The rare radio-loud AGN show jet powers 100-1000 times higher than radio-quiet AGN at the same BH mass and accretion rate, which \citet{Done2026} speculates may be due to a small fraction of AGN harboring high BH spin, though jet power may also depend on the accumulated net magnetic flux rather than spin alone \citep{Tchekhovskoy2011, Sikora2013, Zamaninasab2014, Chamani2021, Savolainen2025arXiv}.

This tension between spin-powered jet and accretion-powered wind is also reflected in the radio-loud/radio-quiet dichotomy. In the spin paradigm, only rapidly spinning BHs produce powerful enough jets to become radio-loud \citep{Rees1982Nat, Wilson1995, Tchekhovskoy+2010, Barausse2012}. However, this requires a wide gap in the spin distribution between the radio-loud and radio-quiet populations. {\sc Mad Sage} naturally produces a broad, continuous spin distribution as a result of the combined action of coherent accretion, incoherent accretion, BH-BH mergers, and BZ spin-down with no imposed gap. This is consistent with the conclusion of \citet{vanVelzenFalcke2013} and \citet{Done2026} that a smooth spin distribution is a generic outcome of BH growth. The radio-loud/radio-quiet dichotomy may instead arise from intermittent jet production or state changes in the accretion disk \citep{vanVelzenFalcke2013}, or from a rare combination of high spin and high magnetic flux \citep{Sikora2013, Chamani2021, Savolainen2025arXiv}. semi-analytic models that include BH spin evolution, such as \citet{Barausse2012} and \citet{Izquierdo-Villalba2020}, similarly find that the spin distribution depends sensitively on the assumed accretion mode and merger history (see also Appendix~\ref{app:BHhistories}).

\subsection{Spin evolution across modelling frameworks}

Several recent works have implemented self-consistent BH spin evolution in cosmological models, providing useful comparisons for {\sc Mad Sage}. \citet{Beckmann2025} tracked BH spin evolution for intermediate-mass and massive BHs in the {\sc NewHorizon} cosmological hydrodynamical simulation, including gas accretion, BH-BH mergers, and BZ-induced jet spin-down. A key finding is that little evidence is found for spin-down through efficient chaotic accretion, in contrast to the predictions of earlier semi-analytic models \citep{Volonteri2005, KingPringle2006}. \citet{Beckmann2025} suggest this may occur because the gas at the BH accretion scale in {\sc NewHorizon} inherits the angular momentum of the surrounding gas at the resolution scale, making the effective accretion more coherent than the isotropic prescription assumed in analytic models. As a result, massive BHs in {\sc NewHorizon} are generally highly spinning, with an average radiative efficiency of $\langle \varepsilon_{\rm thin} \rangle \approx 0.19$. In {\sc Mad Sage}, the incoherent Bondi channel explicitly assigns a random spin orientation at each integration step, more directly encoding the analytic chaotic accretion picture and partially producing the spin-down of massive BHs that {\sc NewHorizon} does not find. Whether this discrepancy reflects a genuine physical difference or a numerical resolution effect remains an open question. 

\citet{Husko2026} presented the {\sc COLIBRE} hybrid AGN feedback model, which implements BH spin evolution in the {\sc swift} code coupled to the full {\sc COLIBRE} cosmological hydrodynamical simulation. Like {\sc Mad Sage}, the model uses the MAD fitting functions of \citet{Ricarte2023Spin}. \citet{Husko2026} find that BH spins are near-maximal for intermediate-mass BHs ($M_{\rm BH} \in [10^6, 10^8]\,\mathrm{M}_\odot$), where efficient thin-disk spin-up dominates over jet spin-down. {\sc Mad Sage} does not reproduce this near-maximal feature. Although BHs within that mass range also assembled most of their mass at high redshift, the incoherent accretion in {\sc Mad Sage} partially counteracts the earlier spin-up. {\sc COLIBRE} has no equivalent incoherent accretion mode, so near-maximal spins acquired during the thin-disk phase are retained at low redshift. 

Beyond this structural difference, BH accretion in {\sc Mad Sage} may be more episodic than in hydrodynamical simulations. Super-Eddington bursts rapidly exhaust the available gas supply, causing $f_{\rm Edd}$ to drop sharply back to sub-Eddington levels rather than tapering smoothly as in {\sc COLIBRE}. Once in the sub-Eddington regime, BHs rarely sustain accretion rates above the critical threshold $f_c$ needed to maintain the thin-disk phase, further limiting the coherent spin-up that would otherwise drive spins toward near-maximal values. At higher masses, BH-BH mergers drive spins toward $|a| \approx 0.7$. In {\sc COLIBRE}, jet feedback dominates over thermal feedback at $z < 0.5$ and $z > 1.5$, with thermal feedback dominating in the intermediate range $0.5 \lesssim z \lesssim 1.5$. Roughly half of the cumulative injected AGN energy is in jet form across all BH masses and redshifts \citep{Husko2026}. 

Lastly, the {\sc L-Galaxies} semi-analytic model \citep{Izquierdo-Villalba2020, Izquierdo-Villalba2022} also tracks BH spin through gas accretion and BH-BH mergers, assuming either coherent or chaotic accretion episodes. Chaotic accretion drives spins toward $|a_\star| \sim 0.1$-$0.3$, consistent with the spin-down of massive BHs found in {\sc Mad Sage}. 

\subsection{Multi-messenger probes of BH spin}
Three complementary observational windows probe BH spin across mass, Eddington ratio, and redshift regimes, each with distinct selection effects that must be taken into account when comparing to model predictions. 

X-ray reflection spectroscopy remains the most direct observational probe for measuring AGN spin, though it carries significant uncertainties \citep{Reynolds2021, Bambi2021}. Spin is inferred by fitting X-ray spectra with reflection models based on the Novikov-Thorne thin disk framework \citep{NovikovThorne1973}, which assumes a geometrically thin, radiatively efficient disk extending all the way to the innermost stable circular orbit. Comparisons between these models and GRMHD simulations of thin disks show broad agreement, but the level of consistency can depend sensitively on disk ionization, magnetic field geometry, and whether intra-ISCO emission is accounted for \citep{Bambi2021, Shashank2022, Shashank2025}.

Beyond these model-specific uncertainties, the thin disk paradigm itself has well known tensions with observations. The standard thin disk predicts optical polarization perpendicular to the jet axis \citep{Schnittman2009}, whereas NASA's Imaging X-ray Polarimetry Explorer \citep{Weisskopf2022} observations of both BH X-ray binaries (stellar mass BHs) and AGN (supermassive BHs) find X-ray polarization parallel to the jet \citep{Krawczynski2022, Gianolli2024}, consistent with a radially extended hot corona rather than a lamppost geometry \citep{Done2026}. Observed AGN disk sizes determined from reverberation mapping and microlensing are also 2-3 times larger than predicted by the standard thin disk temperature profile, and the temperature dependence on luminosity and BH mass is shallower than expected \citep{Hagen2024, Done2026}. 

Beyond X-ray reflection, horizon-scale imaging offers an independent and complementary window onto BH spin that bypasses many of these model assumptions. \citet{Zhang2025ApJngEHT} showed that extensions to the Event Horizon Telescope (EHT) array would grant access to dozens of new SMBH targets beyond M87* and Sgr A*. This would expand horizon-scale spin diagnostics to a population of nearby low-Eddington ratio BHs where ADAF-dominated accretion governs the spin evolution. Current EHT polarimetric data disfavor high-spin configurations for both M87* and Sgr A* based on the polarization angle structure near the horizon \citep{EventHorizonTelescopeCollaboration2021,EHTMW_5_2022}. This is tentatively consistent with the low spins predicted by {\sc Mad Sage} for massive quiescent BHs in the ngEHT/BHEX selection regime.

Finally, LISA will measure BH spin through an entirely different mechanism. The ringdown signal following BH-BH mergers encodes the remnant spin directly in the quasinormal mode frequencies, independent of any accretion model assumptions \citep{Amaro-Seoane2023}. Unlike X-ray or horizon scale measurements, which are sensitive to the current accretion state, the LISA remnant spin carries an integrated record of the entire accretion history of both progenitor BHs, making it a direct probe of the dominant accretion channel across cosmic time. {\sc Mad Sage} \textbf{Full-physics + correction} model predicts remnant spins peaked near $|a_{\star,\rm rem}| \approx 0.7$. A statistical sample of LISA events, in combination with other probes, would allow one to infer the relative contribution of coherent and incoherent accretion to BH growth across a wide range of masses and redshifts. This provides a new multi-messenger probe to constrain the physics of BH feeding that is inaccessible to electromagnetic observations alone.

\section{Conclusions}
\label{S:conclusions}

We have introduced \textsc{Mad Sage}, a semi-analytic galaxy formation model that couples supermassive BH spin evolution to gas accretion and AGN feedback under magnetically arrested disk conditions. By incorporating GRMHD-calibrated spin evolution and jet efficiencies into the {\sc Dark Sage} framework, we explore how the competition between accretion-driven spin-up and electromagnetic spin-down affects both BH populations and their host galaxies. Our results demonstrate that the assumed coherence of BH feeding has important consequences for AGN feedback: incoherent accretion can drive massive BHs toward low spins, weakening electromagnetic jets and reducing the ability of AGN to quench massive galaxies. Accounting for additional mechanical feedback from radiatively inefficient accretion flows substantially improves agreement with the observed stellar mass function. Beyond galaxy evolution, the predicted spin distributions provide testable signatures for X-ray spectroscopy, horizon-scale observations, and future gravitational-wave measurements. Our principal findings are summarized below.

\begin{itemize}
\item The spin models differ most at the high-mass end of the stellar mass function. In the \textbf{Full-physics} model, incoherent accretion drives massive BHs toward the low MAD equilibrium spin, where $\eta_{\rm EM}$ falls far below the \textbf{No-spin} value. AGN feedback is then too weak, and the model overpredicts the number density of galaxies with $M_\star \gtrsim 10^{11}\,{\rm M_\odot}$ for any combination of free parameters. The \textbf{Coherent only} and \textbf{Full-physics + correction} models recover the high-mass end through different mechanisms. In \textbf{Coherent only}, coherent accretion drives BHs to high $|a_\star|$, raising $\varepsilon_{\rm thin}(a_\star)$ and the quasar-mode energy coupled to the cold disk. This also suppresses BH mass growth and gives the weakest radio-mode energy at intermediate stellar masses. In \textbf{Full-physics + correction}, a minimum effective mechanical feedback efficiency \(\eta_{\rm floor}=0.02\), representing additional ADAF outflow channels beyond the electromagnetic BZ jet, provides sufficient halo heating to suppress the overproduction of massive galaxies.
\item All prescriptions predict nearly identical LISA event rates and chirp mass distributions as a result of using the same {\sc TNG100-1} merger trees and halo-mass-dependent BH seeding. The chirp mass distribution is tri-modal, with peaks near $M_{\rm chirp} \sim 10^2\,\mathrm{M}_\odot$ (seed-mass BHs merging before significant growth), $10^3-10^4\,\mathrm{M}_\odot$ (BHs with stagnant or slow growth), and $10^5-10^6\,\mathrm{M}_\odot$ (BHs with sustained super-Eddington accretion at high redshift). The bulk of detectable mergers occurs at $z \lesssim 3$.
\end{itemize}

At low accretion rates where geometrically thick flows are expected, {\sc Mad Sage} may overestimate the magnetic flux threading the horizon. In practice, the transition between MAD and SANE regimes may be gradual rather than sharp, occurring over a range of accretion rates \citep{Narayan2012}. Incorporating a smooth, accretion-rate-dependent magnetization would allow the spin equilibrium to vary continuously between the MAD and SANE limits. This would affect the predicted spin distributions across the full Eddington ratio range probed by our galaxy population.

The isotropic prior on BH--BH merger spin orientations could be replaced by a physically motivated distribution informed by the angular momentum of the host halo. Cosmological simulations suggest that BH spin orientations are not entirely random but retain some memory of the large-scale angular momentum structure of the host \citep{BustamanteVolker2019, Beckmann2024}, which may systematically shift the remnant spin distribution at the high-mass end and affect both the predicted LISA event rates and the AGN energy output at low redshift.

AJPV gratefully acknowledges the support of the National Science Foundation Astronomy \& Astrophysics Postdoctoral Fellowship. This material is based on work supported by the National Science Foundation under Award Number 2502826. AJPV thanks John Forbes for helpful discussions. We used computational facilities from the Yale Center for Research Computing (YCRC). The literature reviews for this work were conducted using NASA’s Astrophysics Data System. P.N., A.R., K.S., H.C., R.N., and B.P. acknowledge support from the Gordon and Betty Moore Foundation and the John Templeton Foundation, which fund the Black Hole Initiative (BHI) at Harvard University. We acknowledge the use of chatgpt for copy-editing this manuscript.

\software{
\textsc{Mad Sage}\footnote{\url{https://github.com/aporrasval/MAD-SAGE}},
Python \citep{VanRossum2009},
NumPy \citep{Harris2020Nat},
SciPy \citep{Virtanen2020},
Matplotlib \citep{Hunter2007},
Astropy \citep{AstropyCollaboration2013, AstropyCollaboration2018, AstropyCollaboration2022},
pandas \citep{McKinney2010, PandasDevelopmentTeam2022}
}

\section*{Data Availability}

The {\sc Mad Sage} source code is publicly available at \url{https://github.com/aporrasval/MAD-SAGE}. The {\sc IllustrisTNG100-1} merger trees and galaxy data used in this work are publicly available at \url{https://www.tng-project.org} \citep{Nelson2019}. The {\sc IllustrisTNG100-1} merger trees were converted from HDF5 to binary format using the converter provided with the \textsc{Sage} semi-analytic model \citep{Croton2016}.\footnote{\url{https://github.com/sage-home/sage-model}} The galaxy catalogues produced in this work will be shared upon reasonable request to the corresponding author.

\bibliography{bridgingscales_bibfile}{}
\bibliographystyle{aasjournal}

\appendix
 
   

This appendix presents supplementary figures supporting the results in the main text. Appendix~\ref{app:AGNenergy_logTSM} shows AGN energy output as a function of stellar mass, and Appendix~\ref{app:BHhistories} traces individual BH evolution histories to illustrate the spin evolution mechanism discussed in Section~\ref{S:results}.

\section{AGN energy impact on galaxy mass}\label{app:AGNenergy_logTSM}

\begin{figure}[t]
    \centering
    \includegraphics[width=\linewidth]{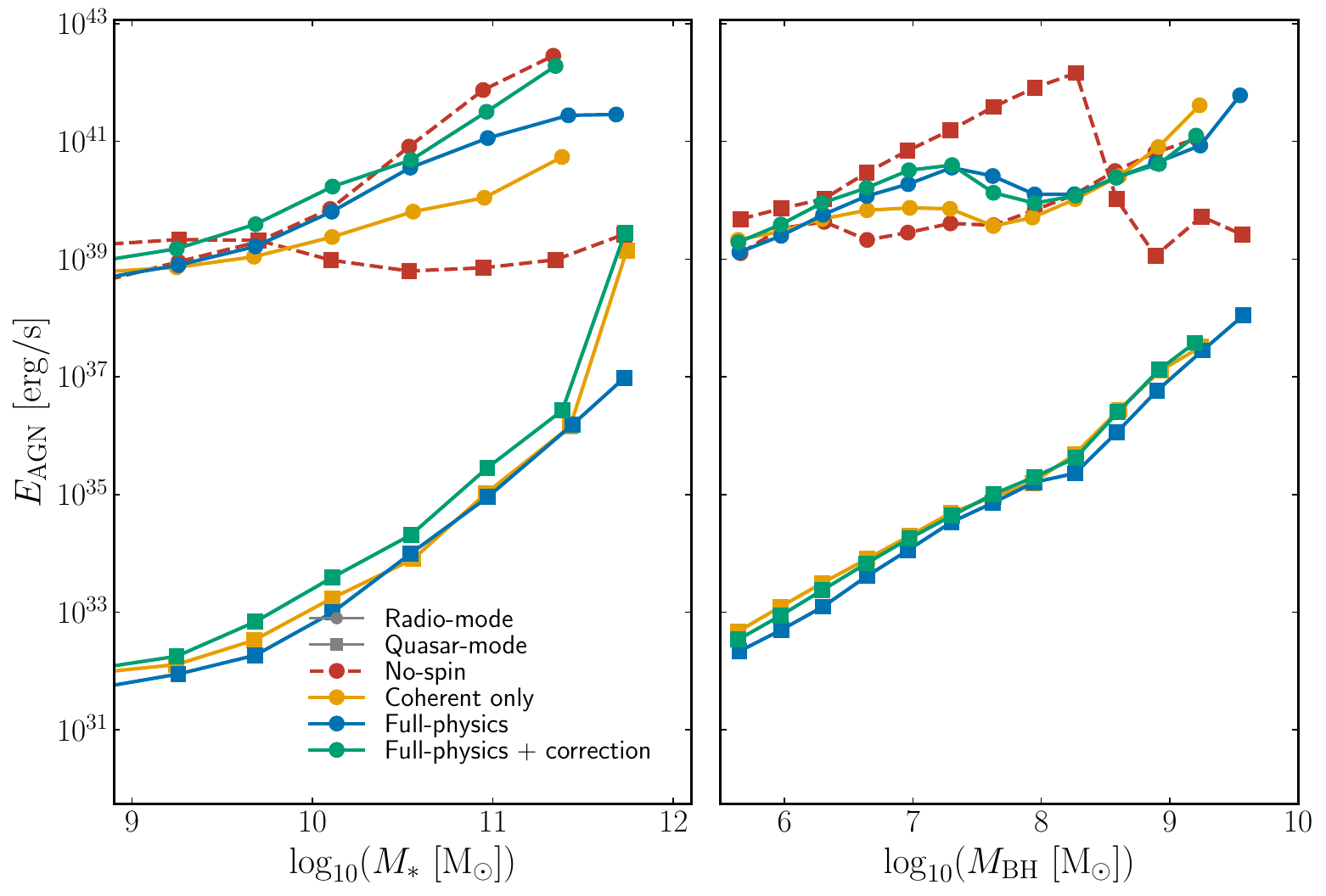}
    \caption{Median \textit{radio-mode} (circles) and \textit{quasar-mode} (squares) AGN energy output as a function of stellar mass (\textit{left}) and BH mass (\textit{right}) at $z=0$ for all models described in Table~\ref{tab:tablemodels}. For all models, the jet power exceeds the \textit{quasar-mode} energy by roughly six to seven orders of magnitude at fixed mass. Among the BH spin models, the \textbf{Coherent only} model produces less jet energy than the \textbf{Full-physics}  and \textbf{Full-physics + correction} models at intermediate stellar masses ($M_* \gtrsim 10^{10}\,\mathrm{M}_\odot$) with the gap closing at the highest BH masses ($M_{\rm BH} \gtrsim 10^8\,\mathrm{M}_\odot$). This suppression arises because the high spins in the \textbf{Coherent only} model raise the radiative efficiency, reducing the fraction of accreted mass retained by the BH. The resulting lower BH masses at fixed stellar mass imply lower Bondi accretion rates and hence less jet energy, despite the higher jet efficiency.
    This suppression at intermediate stellar mass is driven by the lower average BH spin in the \textbf{Coherent only} model for BHs with $M_{\rm BH} \lesssim 10^8\,\mathrm{M}_\odot$. For the \textit{quasar-mode} energy, the \textbf{Coherent only} model closely tracks the \textbf{Full-physics + correction} model across both stellar mass and BH mass.}
    \label{fig:AGNfeedback_logTSM_logBHM}
\end{figure}

Figure~\ref{fig:AGNfeedback_logTSM_logBHM} shows the median \textit{radio-mode} and \textit{quasar-mode} AGN energy output as a function of stellar mass and BH mass at $z=0$ for all models. At low stellar masses, the \textbf{No-spin} model produces slightly less radio-mode energy than the \textbf{Full-physics + correction} model, while at high stellar masses, the \textbf{No-spin} radio-mode power becomes slightly higher. The behavior as a function of BH mass is markedly different. The \textbf{No-spin} model has the highest radio-mode power of all models up to $M_{\rm BH} \sim 10^8\,\mathrm{M}_\odot$, after which it declines sharply by roughly three orders of magnitude and becomes the least powerful mode at the highest BH masses.

Among the BH spin models, the \textbf{Coherent only} model produces the least \textit{radio-mode} energy at intermediate stellar masses. Coherent accretion drives BHs towards high spin, which raises the Novikov-Thorne radiative efficiency $\varepsilon_{\rm thin}(a_\star)$ and reduces the fraction of accreted mass retained by the BH. This results in lower average BH masses at fixed stellar mass, lower Bondi accretion rates, and therefore, less cumulative \textit{radio-mode} heating relative to the \textbf{Full-physics} model, where chaotic accretion episodes reduce the average spin and the associated radiative losses. The gap narrows at the highest BH masses ($M_{\rm BH} \gtrsim 10^8\,\mathrm{M}_\odot$), where the elevated spin in the \textbf{Coherent only} model boosts the BZ efficiency. 

The \textbf{Full-physics + correction} model most closely tracks the \textbf{No-spin} \textit{radio-mode} energy across the full stellar mass range, consistent with it providing the best agreement with the observed stellar mass function. The mechanical feedback floor ensures a minimum level of halo heating even at low spin, partially compensating for the reduced BZ efficiency and bringing the total feedback energy closer to the \textbf{No-spin} level.

For the \textit{quasar-mode} energy, the \textbf{Coherent only} model closely traces the \textbf{Full-physics + correction} model across both stellar mass and BH mass. The quenching differences seen at the high-mass end of the stellar mass function (Figure~\ref{fig:SMF_z0}) therefore arise from a combination of the \textit{radio-mode} differences between spin models, the large scatter in BH mass at fixed stellar mass at the high mass end of the scaling relation, and the upturn in \textit{quasar-mode} energy at high stellar masses in the \textbf{Coherent only} and \textbf{Full-physics + correction} models.

\section{Black hole mass, spin, and accretion histories at the individual level}\label{app:BHhistories}

\begin{figure*}
    \centering
    \includegraphics[width=\linewidth]{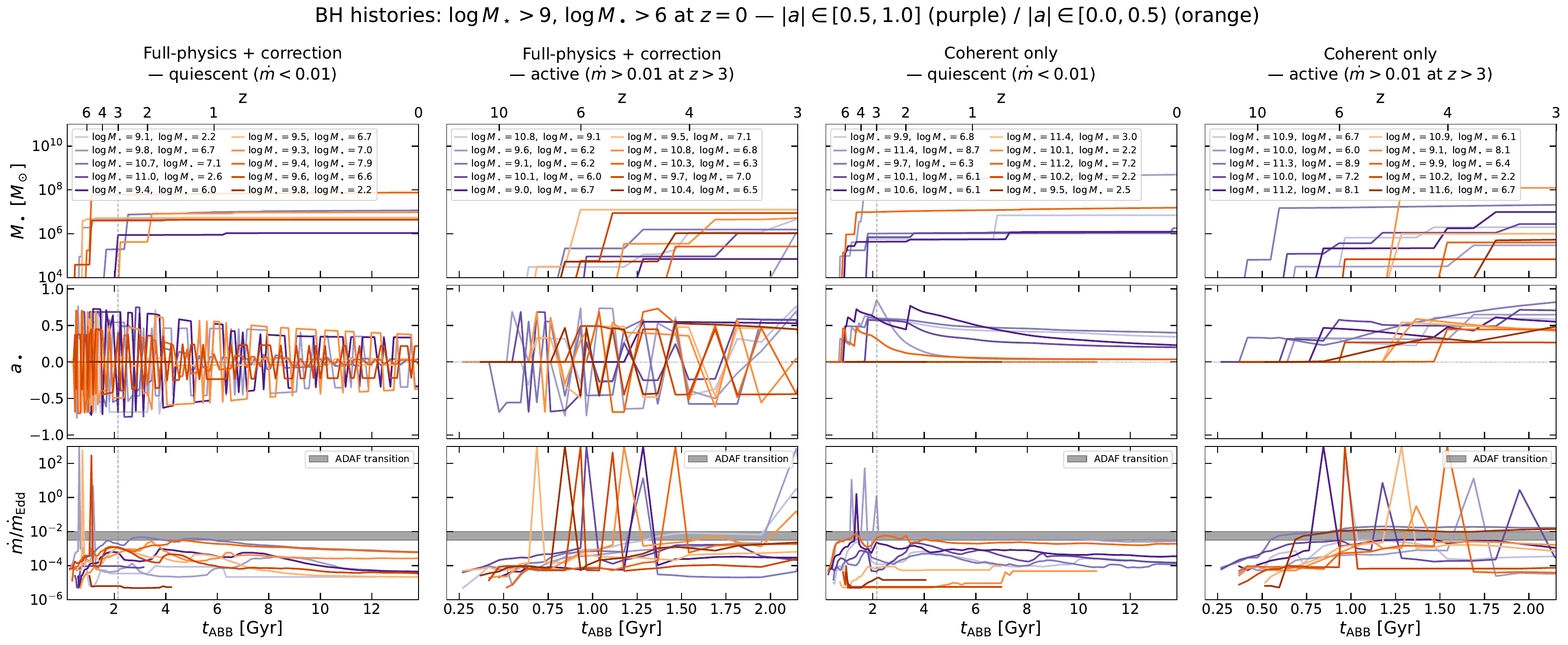}
    \caption{Sample of BH evolution histories for central galaxies at $z=0$ with $\log(M_\star/\mathrm{M}_\odot) > 9$ and $\log(M_\bullet/\mathrm{M}_\odot) > 6$, comparing the \textbf{Full-physics + correction} (left two columns, also labeled Full Physics) and \textbf{Coherent only} (right two columns) models. Within each model, BHs are split into \emph{quiescent} systems (those with $f_{\rm Edd} < 0.01$ at all snapshots below $z < 3$) and \emph{active} systems that sustained at least one episode with $f_{\rm Edd} > 0.01$ above $z > 3$. The classification boundary is marked by the vertical dashed line. The three rows show, from top to bottom, the BH mass $M_\bullet$, the spin parameter $a_\star$, and the Eddington ratio $f_{\rm Edd}$ as a function of cosmic time. Purple (orange) tracks correspond to BHs whose spin magnitude satisfies $|a_\star| \in [0.5, 1.0]$ ($|a_\star| \in [0, 0.5)$) at $z\sim3$. The grey shaded band in the lower panels indicates the ADAF transition region around $f_{\rm Edd} \approx 0.01$. Note that active columns span only $t_{\rm ABB} \lesssim 2\,\mathrm{Gyr}$ ($z \gtrsim 3$), where most BHs are accreting near or above the Eddington rate. The quiescent columns show the full subsequent evolution to $z=0$.}
    \label{fig:BHhistories}
\end{figure*}

Figure~\ref{fig:BHhistories} displays a striking contrast between spin evolution models. In the \textbf{Full-physics + correction} model, the spin tracks of quiescent BHs are highly erratic at late times ($z < 3$). Repeated episodes of low angular momentum Bondi accretion from the hot halo flip $a_\star$ between prograde and retrograde values on timescales of $\lesssim 200$ Myr. This prevents any coherent spin-up. As a result, most quiescent BHs in this model end up in the low-spin range, regardless of their mass. By contrast, the \textbf{Coherent only} model produces smooth, monotonically evolving spin histories. Because most BHs at high-z ($z>3$) accrete near the Eddington limit or above, there is little-to-no Bondi accretion, and thus, no spin-flip mechanism at low $f_{\rm Edd}$. As a result, $a_\star$ drifts slowly towards intermediate or high values and largely stays there. Depending on how early these episodes of accretion occur, some BHs may experience spin-down due to BZ effects. 

The high-z active panels ($z>3$) are qualitatively more similar between models. At high accretion rates, both models channel mass through the thin disk and spin evolution is dominated by coherent Lense-Thirring alignment. The broader spread in spin amplitude for the active BHs in the \textbf{Full-physics + correction} model reflects occasional incoherent episodes interspersed within the burst. The BH mass histories are nearly indistinguishable between models for approximately similar BH mass. This is expected given that the two prescriptions differ only in how angular momentum is assigned to infalling gas, not in the total accreted mass. Together, the panels show that the late-time spin distribution is set overwhelmingly by the mode of accretion at $z<3$ rather than by the high-z assembly history.

\end{CJK}
\end{document}